\documentclass[a4paper,11pt]{article}
\usepackage{jheppub} % for details on the use of the package, please
\usepackage{slashed}
\usepackage[]{mdframed}
\usepackage[T1]{fontenc} % if needed

\newcommand{\nn}{\nonumber}

\newcommand{\bra}[1]{\mbox{$\langle #1 |$}}
\newcommand{\ket}[1]{\mbox{$| #1 \rangle$}}

\newcommand{\Tr}{{\rm Tr}\,}

\newcommand{\be}{\begin{equation}}
\newcommand{\ee}{\end{equation}}
\newcommand{\bea}{\begin{eqnarray}}
\newcommand{\eea}{\end{eqnarray}}

\title{\boldmath 
Spread Complexity for Local Operator Quenches with Conserved Momentum}
\author[a]{Ali Fatemiabhari}

\affiliation[a]{Institute for Theoretical and Mathematical Physics, Lomonosov Moscow State University, 119991 Moscow, Russia}
\emailAdd{alifatemiabhari@gmail.com}
\abstract{We study a family of locally excited states in two-dimensional conformal field theories in different setups, defined on the plane, on the cylinder, or at finite temperature. These states are constructed by assigning independent chiral and anti-chiral regulators to a local primary operator, leading to excitations carrying net momentum. 
We analytically calculate the first few Lanczos coefficients and the early time behaviour of the spread complexity associated with these states. 
Then, using numerical methods, we obtain the coefficients reaching larger Krylov indices and find the associated spread complexity for a longer range of time evolution.
We also compare the rate of change of the resulting spread complexity with the proposed holographic dual quantity, related to the proper momentum of an infalling particle in the dual bulk, but we observe a mismatch.
}
\begin{document} 
\maketitle
\flushbottom
%\newpage
%%%%%%%%%%%%%%%%%%%%%%%%%%%%%%%%%%%%%%%%%%%%%%%%%%%%%%%
\section{Introduction}\label{introgeneral}
The growth of complexity under unitary time evolution has become a unifying theme across different disciplines such as quantum information, many-body chaos, and holography.
Of the many proposals for making the definition of complexity precise, those based on Krylov subspaces are distinguished by being canonical: once a Hamiltonian and an initial state or operator are specified, no further input, such as gate set, tolerance, or cost function, is required. This is in contrast with circuit
complexity, and the construction is algorithmic. Repeated action of $H$ on a seed
state, or of the Liouvillian $\mathcal{L}=[H,\,\cdot\,]$ on a seed operator, followed by Gram-Schmidt orthonormalisation, generates through the Lanczos
algorithm an orthonormal Krylov basis $\{|K_n\rangle\}$. In this basis, the generator of time evolution is tridiagonal; Hence, the dynamics reduces to that of a single particle on a one-dimensional chain, with on-site energies $a_n$ and
nearest-neighbour hopping amplitudes $b_n$ \cite{Parker:2018yvk}.

The application of this construction to operators defines Krylov complexity, the mean position of the particle on the chain. Its long-time behaviour is controlled by the large-$n$ asymptotics of the Lanczos coefficients. In chaotic systems, $b_n$ grows asymptotically linearly, $b_n\simeq\alpha n$, and the slope $\alpha$ bounds the Lyapunov exponent at infinite temperature, $\lambda_L\leq2\alpha$ \cite{Parker:2018yvk}. Applying the process to states leads to spread complexity \cite{Balasubramanian:2022tpr,Alishahiha:2022anw}. Here, the Krylov basis is unique, as among all bases containing the initial state, it minimises the spread of the time-evolved wavefunction over a finite interval of time. Spread complexity is thus intrinsic to the pair $(H,|\psi(0)\rangle)$, and can be computed directly from the autocorrelation function, defined later in the text.

Much of the appeal of these measures stems from the fact that the entire dynamics is compressed into a single sequence of real numbers $\{a_n,b_n\}$. This has enabled distinguishing chaotic from
integrable dynamics \cite{Balasubramanian:2022tpr,Rabinovici:2023yex}, extensions to quantum field theory (QFT) \cite{Avdoshkin:2022xuw,Caputa:2024xkp}, and connections to geometric and circuit formulations of complexity \cite{Caputa:2021sib,Muck:2026top}, making this definition of complexity a multi-purpose observable. We refer the reader to \cite{Nandy:2024evd,Rabinovici:2025otw,Baiguera:2025dkc,Jeong:2026gdc} for recent reviews.
%%%%%%%
\subsection{Spread/Krylov complexity in QFTs}
%%%%%
Computing Krylov and spread complexity in QFT is considerably more subtle than in finite-dimensional quantum systems, both conceptually and technically. The Lanczos coefficients are determined by the full set of moments of the autocorrelation function, or equivalently by its power spectrum, and in a QFT, the power spectrum generically has unbounded support. As a consequence, the large-$n$ behaviour of $b_n$ is dictated by the analytic structure of the
correlator, not by the dynamics of the theory \cite{Dymarsky:2021bjq,Vasli:2023syq,Alishahiha:2024vbf,Imani:2025etp,Alishahiha:2026fnu}. For instance, in any thermal field theory, the autocorrelation function is singular at $\tau=\beta/2$. This forces linear growth of $b_n$ coefficients with slope fixed by $\pi/\beta$ value and hence exponential growth of Krylov complexity even in free and rational conformal field theories (CFT)s are not a sign of chaos \cite{Camargo:2022rnt,He:2024xjp,Baume:2026jyt}.

%The operator growth hypothesis \cite{Parker:2018yvk} therefore holds trivially
%in field theory and ceases to be a diagnostic of chaos. 
% A systematic analysis of free, weakly interacting and holographic models \cite{Avdoshkin:2022xuw}
% showed that any relation between the growth of $b_n$ and chaos can at best hold
% in a truly asymptotic regime controlled by UV physics, and that Krylov complexity in QFT can behave qualitatively differently from its holographic counterparts. One way to restore sensitivity to the dynamics is to impose a bounded power spectrum, through a UV cutoff or a lattice regularisation, as done for free and interacting scalar theories in \cite{Camargo:2022rnt,He:2024xjp} with results being regulator-dependent, and the extension to general temperatures already requires truncation schemes for the Wightman power spectrum \cite{He:2024xjp}. 
Closed-form results are almost only available when the dynamics is organised by a symmetry algebra, such as $sl(2,\mathbb{R})$ and Heisenberg-Weyl structures underlying coherent-state evolution \cite{Caputa:2021sib,Balasubramanian:2022tpr},
emergent Poincare symmetries in operator growth \cite{Magan:2020iac}, or Virasoro dynamics in 2d CFTs \cite{Caputa:2021ori}, where the answer is moreover sensitive to the choice of state \cite{Kundu:2023hbk}. Without these symmetries, one should solve the moment problem directly, either through the recursion method or through the Toda-chain formulation \cite{Dymarsky:2019elm}, and both routes are numerically delicate. As explained later in the appendices, the Hankel determinants encoding the moments become exponentially ill-conditioned, while the Lanczos algorithm itself suffers from loss of orthogonality and requires high-precision arithmetic to reach to large Krylov indices \cite{Rabinovici:2020ryf}.
%%%%

In QFTs, understanding the dynamics far from equilibrium is of utmost importance \cite{Eisert:2014jea,Balasubramanian:2010ce}. The interplay between quenches, information transport, and the effect of conserved charges in the propagation of local excitations has absorbed vast attention \cite{Srednicki:1994mfb,DAlessio:2015qtq}.
Specifically, 2d CFTs are interesting and simpler platforms for deriving universal conclusions about the dynamics of quantum quenches and evolution of observables in their presence \cite{Bernard:2016nci,Calabrese:2016xau}. In \cite{David:2026owc, Caputa:2026hxt}, authors introduce a certain family of locally excited states. These states are constructed by assigning independent chiral and anti-chiral regulators to local primary operators. The asymmetric regulator leads to local operator quenches that carry conserved longitudinal or angular momentum. This provides a great opportunity by using the symmetries of a 2d CFT to obtain the autocorrelation function of the excited states exactly. 

The main goal of this paper is to calculate the Lanczos coefficients and spread complexity associated with these states. We do so analytically for the first few Lanczos coefficients and the early-time behaviour of the spread complexity. Using numerical methods with high-precision arithmetic, we obtain these coefficients reaching larger Krylov index numbers and then find the associated spread complexity for a range of time evolution.

%%%%
It is interesting that the local operator excitations in the field theory have a description in a dual gravity setup given by massive particles that propagate in AdS background and using the duality, many observables can be calculated \cite{Nozaki:2013wia,Caputa:2014vaa,Asplund:2011cq,Asplund:2013zba,Asplund:2014coa,Caputa:2015tua,David:2016pzn,Bhattacharyya:2019ifi,Berenstein:2019tcs,Mao:2024cnm,Mao:2025cfl,Mao:2025hkp,Kudler-Flam:2023ahk}. This duality is obtained based on the gauge/gravity correspondence \cite{Maldacena:1997re}. In \cite{David:2026owc,Caputa:2026hxt} the authors also study the holographic dictionary that relates point particles in asymptotically AdS$_3$ spacetimes that carry longitudinal or angular momentum to states discussed earlier with independent chiral and anti-chiral regulators. Hence, we make use of this dictionary to compare the calculated spread complexity with dual holographic observables that are proposed to be dual to the spread complexity.
%%%%%%%
\subsection{Holographic duals}
 The holographic dual to spread/Krylov complexity originates from the idea that the growth of the complexity in a boundary is related to the momentum of an object falling into the bulk \cite{Susskind:2018tei,Susskind:2019ddc,Susskind:2020gnl,Barbon:2020uux}. An explicit link to Krylov dynamics first appeared in Jackiw-Teitelboim (JT) gravity \cite{Rabinovici:2023yex}, and \cite{Caputa:2024sux} proposed identifying the rate of spread complexity with the proper radial momentum of a massive probe particle. This relation has since been inspected and generalised beyond the $sl(2,\mathbb R)$ and JT settings, in many different directions \cite{Fan:2024iop,He:2024pox,Heller:2024ldz,Das:2024tnw,Ambrosini:2024sre,Fu:2025kkh,Jeong:2026iac,Li:2025observer,Li:2026comments,Muck:2026top,Qu:2025lanczos,Aguilar-Gutierrez:2025kmw,Alfinito:2026vah,Qu:2026dmv}.

The extension of \cite{Caputa:2024sux} from 2d CFTs to $\mathcal N=4$ SYM is given in \cite{Fatemiabhari:2025cyy}  and in confining backgrounds such as the Anabal\'on--Ross soliton \cite{Anabalon:2021tua}, complexity exhibits oscillations whose frequency is set by the confinement scale \cite{Fatemiabhari:2025usn,Fatemiabhari:2026goj}, a feature mirrored in the Ising chain \cite{Jiang:2025wpj} and traced more generally to the presence of a discrete spectrum \cite{Nunez:2026vhw}. In conformal quiver theories in four and six dimensions, motion along internal directions shapes the early-time growth while the universal late-time behaviour is preserved \cite{Fatemiabhari:2025poq,Fatemiabhari:2026six}. Charged, composite and extended probes were studied in \cite{Nastase:2026lhz}, and the acceleration of spread complexity was related to holographic $c$-functions in \cite{Nunez:2026rgflow}. These results have led to diverse tests and extensions to other setups \cite{Roychowdhury:2026eta,Zoakos:2026coulomb,Roychowdhury:2026lin,Roychowdhury:2026bmnstate,Roychowdhury:2026planewave,Roychowdhury:2026bmnspectral,Fadafan:2026lifshitz}. 
%%%%%%

We will concentrate on the new proposals provided in \cite{Chatzis:2026ekd,Chatzis:2026oou} to compute the holographic spread complexity and compare it with our CFT results. This proposal generalises the original duality of rate of change of complexity dual to the proper momentum of a particle to probes carrying conserved momentum.  

The rest of the paper is organised as follows. In Section~\ref{sec:spread}, we review the notion of spread complexity and provide a general methodology to calculate Lanczos coefficients and the complexity itself. Section~\ref{sec:CFTKry} provides a review of the locally excited states discussed above and computes their associated spread complexity. In Section~\ref{sec:Holography}, we investigate the holographic duals to the states studied in Section~\ref{sec:CFTKry} and compare CFT spread complexity results with the holographic dual proposals. Section~\ref{conclsect} contains conclusions and outlook. We leave many technical details about analytic and numerical calculations of Lanczos coefficients and spread complexity to Appendices~\ref{app:Lanc1} and \ref{app:Lanc}. Finally, Appendix~\ref{app:gravity} provides more details about dual backgrounds to the local quenches.

\section{Review of spread complexity} \label{sec:spread}
We begin by reviewing the notion of spread complexity to quantify the complexity of an evolving state in terms of its spread over the Hilbert space. A pedagogical introduction and further technical details can be found in \cite{Balasubramanian:2022tpr, Nandy:2024evd,Rabinovici:2025otw,Jeong:2026gdc}.

Consider the unitary time evolution of a state,
\begin{equation}
\label{eq:psit}
|\psi(t)\rangle=e^{-\mathrm{i} H t}|\psi(0)\rangle\,.
\end{equation}
Spread complexity was introduced to quantify the complexity associated with this evolution. 
One can start from the ordered set of vectors
\begin{equation}
\label{eq:Hn}
\left\{ H^n |\psi(0)\rangle \; \middle| \; n = 0, 1, 2, \ldots \right\}\,,
\end{equation}
and constructs the so-called Krylov basis given as
\begin{equation}
\label{eq:krylovBasis}
\mathcal{K} = \left\{ \left| K_n \right\rangle, \; n = 0, 1, 2, \ldots, |\mathcal{K}|-1 \right\}\,,
\end{equation}
by Gram-Schmidt (GS) orthonormalisation of states in Eq.~\eqref{eq:Hn}, \cite{Lanczos:1950zz}.
This process, called the Lanczos algorithm, yields the basis in which $H$ acts tridiagonally. Explicitly, setting $\ket{K_0}=\ket{\psi(0)}$, the basis is generated recursively through
\begin{equation}
\label{eq:Krylovbasis}
H\left|K_n\right\rangle=a_n\left|K_n\right\rangle+b_n\left|K_{n-1}\right\rangle+b_{n+1}\left|K_{n+1}\right\rangle\,,
\end{equation}
where the Lanczos coefficients $a_n$ and $b_n$ are fixed by
\begin{equation}
a_n=\langle K_n|H|K_n\rangle\,,\qquad
b_{n+1}=\langle A_{n+1}|A_{n+1}\rangle^{1/2}\,,\qquad
\ket{A_{n+1}}\equiv\left(H-a_n\right)\ket{K_n}-b_n\ket{K_{n-1}}\,,
\label{eq:lanczosCoeff}
\end{equation}
together with $\ket{K_{n+1}}=b_{n+1}^{-1}\ket{A_{n+1}}$. The recursion terminates at $n=|\mathcal{K}|-1$, where $b_{|\mathcal{K}|}=0$. Expanding the time-evolved state in this basis,
\begin{equation}
|\psi(t)\rangle = \sum_{n=0}^{|\mathcal{K}|-1} \phi_n(t) | K_n \rangle, \qquad
\phi_n(t) \equiv \langle K_n | \psi(t) \rangle\,,
\end{equation}
one finds that the amplitudes $\phi_n(t)$, whose moduli squared define the probabilities $p_n(t)=|\phi_n(t)|^2$, obey the discrete Schr\"odinger equation
\begin{equation}
\label{eq:schrEq}
\mathrm{i} \partial_t \phi_n(t)=a_n \phi_n(t)+b_{n+1} \phi_{n+1}(t)+b_n \phi_{n-1}(t)\, .
\end{equation}
Equation \eqref{eq:schrEq} provides a transparent interpretation for the dynamics: the evolution in the Krylov basis is equivalent to that of a single quantum particle hopping on a semi-infinite one-dimensional lattice,  so-called
\textit{Krylov chain}. On this chain, on-site energies are given by $a_n$ and nearest-neighbour
hopping amplitudes are $b_n$. Since site $n$ is in one-to-one correspondence with the Krylov vector $\ket{K_n}$, the mean position of the particle measures how far the state has spread across the Krylov basis. Accordingly, spread complexity is defined as the expectation value of the position operator on the
Krylov chain \cite{Balasubramanian:2022tpr},
\begin{equation}
\label{eq:Ctdef}
C(t)=\langle n\rangle = \sum_{n=0}^{|\mathcal{K}|-1} n\left| \phi_n(t) \right|^2\,.
\end{equation}
Knowledge of the Lanczos coefficients is therefore a prerequisite for
evaluating the amplitudes $\phi_n(t)$ and, in turn, the spread complexity.
This data can be obtained from the return amplitude
\begin{align}
\label{eq:Stdef}
S(t) &\equiv\langle\psi(t) \mid \psi(0)\rangle=\langle\psi(0)| e^{\mathrm{i} H t}|\psi(0)\rangle=\phi_0^*(t)\,,\\
&\left.\mu_n \equiv \frac{d^n}{d t^n} S(t)\right|_{t=0}=\left\langle K_0\right|(\mathrm{i} H)^n\left|K_0\right\rangle\,,
\end{align}
by means of the moment recursion method \cite{Viswanath1994TheRM}. Specifically, the
Lanczos coefficients are encoded in the moments of the return amplitude,
in the sense that each $\mu_n$ is a polynomial in $\{a_m,b_m\}_{m\leq n}$; these
relations can be inverted iteratively to express the Lanczos coefficients in
terms of the moments. Appendix~\ref{app:Lanc1} provides some more details for the derivation of the Lanczos coefficients from the return amplitude. 

At early times, the dependence of the spread complexity on Lanczos coefficients can be computed explicitly and is controlled by the first few coefficients only. The leading orders of the early-time expansion read
\cite{Fan:2022xaa}
\begin{equation}
C(t) = b_1^2 t^2 + \left( \frac{1}{6} b_1^2 b_2^2 - \frac{1}{3} b_1^4 - \frac{1}{12} (a_0 - a_1)^2 b_1^2 \right) t^4 + \mathcal{O}(t^6)\,.
\label{eq:earlyCt}
\end{equation}
This is dictated by unitarity and is a universal result irrespective of specific system details. We will make use of this general result in the analysis below. 

Now we will apply the general method described above to certain states in 2d CFTs to gain knowledge about their time evolution and complexity.

\section{Spread complexity for holographic local operator quenches}\label{sec:CFTKry}
In this section, we review a universal family of quantum states in 2d CFTs, constructed by assigning independent chiral and anti-chiral regulators to a local primary operator, introduced in \cite{David:2026owc,Caputa:2026hxt}. The physical interpretation of these states is local operator quenches carrying finite momentum, generalising the standard local quench setup, in which the two regulators are equal. Upon an explicit identification of bulk and boundary parameters, it has been shown that these states reproduce the same stress tensor in CFT as the holographic stress tensors in each of their dual backgrounds.

These states provide concrete realisations of local operator excitations in 2d CFTs, evolved in real time. We first review their definition and the stress-tensor one-point functions, which the operator product expansion (OPE) fixes, and then compute the spread complexity for each case. The calculation of return amplitude and some of the Lanczos coefficients are derived analytically, while the long-term evolution for the spread complexity is computed numerically.
%%%%%%%%%%
\subsection{CFT on the plane} \label{sec:CFTLin}
%%%%%%%%%%
In this section, we work in a 2d CFT in Euclidean signature on the complex plane, with
coordinates $(z,\bar{z})=(x+i\tau,\,x-i\tau)$, and consider the family of states
\begin{align}
\ket{\psi(t)}\equiv e^{-itH}\,e^{-\frac{\epsilon_++\epsilon_-}{2}H}\,
e^{i\frac{\epsilon_+-\epsilon_-}{2}P}\,{\cal O}_\Delta(0,0)\ket{0}\,,
\end{align}
where $\ket{0}$ is the CFT vacuum and ${\cal O}_\Delta$ is a spinless primary of
conformal dimension $\Delta=h+\bar{h}=2h$, inserted at the origin
$z=\bar{z}=0$. The corresponding density matrix is
\begin{equation}
\rho(t)=\mathcal{N}\ket{\psi(t)}\bra{\psi(t)}
\equiv\mathcal{N}\,{\cal O}_\Delta(z_2,\bar{z}_2)\ket{0}\bra{0}{\cal O}_\Delta^\dagger(z_1,\bar{z}_1)\,,
\label{eq:rhotLin}
\end{equation}
with $\mathcal{N}$ fixed by $\Tr\rho(t)=1$ and with insertion points
\begin{eqnarray}
&&z_2=-i(\epsilon_-+it)\,,\qquad z_1=i(\epsilon_--it)\,,\nn\\
&&\bar{z}_2=i(\epsilon_++it)\,,\qquad \bar{z}_1=-i(\epsilon_+-it)\,.
\label{eq:z12Lin}
\end{eqnarray}
The chiral and anti-chiral regulators, therefore, control the holomorphic and
anti-holomorphic insertion points separately. 

The generators $H$ and $P$ implement translations in Euclidean
time $\tau$ and in space $x$ respectively, and both annihilate the vacuum.
The state is thus a local excitation of the vacuum, smeared in Euclidean time
and in space by amounts controlled independently by $\epsilon_+$ and
$\epsilon_-$, and subsequently evolved in real Lorentzian time $t$. Following the standard treatment of quantum quenches in 2d CFTs
\cite{Calabrese:2016xau}, we carry out all computations treating $\tau$ coordinate arguments as real numbers. Hence the expression  
$(\epsilon_++\epsilon_-)/2+it$ is taken as a real Euclidean time, and continued
to real $t$ only at the end. Note that, after this continuation, $z$ and
$\bar{z}$ are independent variables rather than complex conjugates. For
$\epsilon_+=\epsilon_-$ the construction reduces to the standard local operator
quench \cite{Nozaki:2014hna,He:2014mwa,Caputa:2014vaa}; the novelty here is
that $\epsilon_+\neq\epsilon_-$ promotes the excitation to carry finite longitudinal
momentum.

% At $t=0$ this reads
% \begin{equation}\label{eq:rho0Lin}
% \rho(0)= \mathcal{N}\,{\cal O}(-i\epsilon_-,i\epsilon_+)\ket{0}\bra{0}
% O^\dagger(i\epsilon_-,-i\epsilon_+)\,.
% \end{equation}

The stress-tensor one-point functions in this state follow from the universal OPE and are therefore independent of any dynamical data beyond
$\Delta$. Writing $x_\pm=x\pm t$, one finds
\begin{equation}
\langle T_{\mp\mp}(x_\mp)\rangle
=\frac{\Delta}{\pi}\,\frac{\epsilon_\mp^{2}}{\left(x_\mp^{2}+\epsilon_\mp^{2}\right)^{2}}\,,
\label{TpmPoincare}
\end{equation}
so that each chirality is regulated by its own parameter. It was shown that this stress tensor matches the one obtained from the holographic background dual to this state--- see Ref.~\cite{Caputa:2026hxt}.

Integrating over the spatial slice gives
the conserved energy and momentum
\begin{equation}
E=\int dx\, T_{tt}=\int dx \left(T_{++}+T_{--}\right)=\frac{h}{\epsilon_-}+\frac{h}{\epsilon_+}=\frac{h}{\epsilon}(\tilde \epsilon+\tilde \epsilon^{-1})\,,
\label{eq:linEn}
\end{equation}
\begin{equation}
P=\int dx\, T_{tx}=\int dx \left(T_{--}-T_{++}\right)=\frac{h}{\epsilon_-}-\frac{h}{\epsilon_+}=\frac{h}{\epsilon}(\tilde \epsilon-\tilde \epsilon^{-1})\,,
\label{eq:linMo}
\end{equation}
where we have introduced the new parameters 
\begin{equation}
\epsilon\equiv\sqrt{\epsilon_+\epsilon_-}, \quad \tilde \epsilon\equiv\sqrt{\frac{\epsilon_+}{\epsilon_-}},%=R\,e^{\eta_1}\,,
\label{eq:Linetaepm}
\end{equation}
for later convenience and making a connection with the holographic results later.
Since $E^2-P^2=(\frac{\Delta}{\epsilon})^2$ is independent of $\tilde\epsilon$, the second
parameter acts precisely as a boost rapidity: $\epsilon$ sets the overall
smearing scale, and hence the rest energy of the excitation, while $\tilde{\epsilon}$
boosts it along the spatial direction. 

The dynamics of this state is as follows.
In the limit $\epsilon_+=\epsilon_-=\epsilon$, the state describes an entangled
pair of excitations that is propagating to the right and the left of the origin at the speed of light \cite{Nozaki:2013wia}. The energy of the state will be given by $\frac{\Delta}{\epsilon}$ and smearing is set by the parameter $\epsilon$. 
In the generic case, two excitations will have different peak height for left-moving and right-moving energy density pulses \cite{Caputa:2026hxt}. Parameter $\epsilon=\sqrt{\epsilon_+\epsilon_-}$ still sets the overall smearing scale while the ratio $\sqrt{\epsilon_+/\epsilon_-}$ determines the net linear momentum. Overall, smaller $\epsilon$ values lead to less smearing of states in the Hilbert space of the theory. 

\textbf{Spread complexity:} Now, we focus on the calculation of the spread complexity of the associated state
\begin{equation}
\ket{\psi(t)}=e^{-iHt}\ket{\psi_0}, \quad \ket{\psi_0}=\,e^{-\frac{\epsilon_++\epsilon_-}{2}H}\,
e^{i\frac{\epsilon_+-\epsilon_-}{2}P}\,O(0,0)\ket{0}.
\end{equation}
Following the procedure described in Section~\ref{sec:spread}, we start the computation from the return amplitude \cite{Caputa:2023vyr},
\begin{equation}
S(t)^*=\text{Tr}(\rho(0)e^{-iHt})=\frac{\langle \mathcal{O}^\dagger(z_3,\bar{z}_3)\mathcal{O}(z_4(t),\bar{z}_4(t))\rangle}{\langle\mathcal{O}^\dagger(z_3,\bar{z}_3) \mathcal{O}(z_4(0),\bar{z}_4(0))\rangle},
\end{equation}
where the insertion points are derived from Eqs.(\ref{eq:rhotLin}) and (\ref{eq:z12Lin}),
\begin{align}
z_3=i\epsilon_-,&\quad \bar{z}_3=-i\epsilon_+,\nonumber\\
z_4(t)=-i(\epsilon_- +it)&,\quad \bar{z}_4(t)=+i(\epsilon_+ +it).  
\end{align}
The two-point correlator functions can be calculated using the general results for a CFT on a plane 
\begin{equation} \label{eq:propLin}
\langle\mathcal{O}^\dagger(z_1,\bar{z}_1) \mathcal{O}(z_2,\bar{z}_2)\rangle=z^{-2h}_{12}\bar{z}^{-2\bar{h}}_{12},
\end{equation}
with $z_{ij}=z_i -z_j$. In this case, the return amplitude takes the simple form
\begin{equation}
S(t)=\left(1-\frac{it}{2\epsilon_+}\right)^{-\Delta}\left(1-\frac{it}{2\epsilon_-}\right)^{-\Delta}.\label{eq:StLin}
\end{equation}

%%%%%%%%%%%

Using this autocorrelation function, one can derive the moments from Eq.~\eqref{eq:Stdef} that are used to find the Lanczos coefficients. We use the Toda hierarchy technique as described in Appendix~\ref{app:Lanc1} to find the analytic expressions for the Lanczos coefficients. Here we provide the first few coefficients with the convention $b_0=0$,
\begin{align} \label{eq:anbnLin1}
   a_0&=\frac{1}{2}   \left(\frac{1}{\epsilon_-}+\frac{1}{\epsilon_+}\right)\Delta,\\
   b_1&=\frac{1}{2} \sqrt{\Delta  \left(\frac{1}{\epsilon_-^2}+\frac{1}{\epsilon_+^2}\right)},\\
   a_1&=\frac{1}{2}   \left(\frac{1}{\epsilon_-}+\frac{1}{\epsilon_+}\right)\left((\Delta+1)+\frac{(\epsilon_--\epsilon_+)^2}{\epsilon_+^2+\epsilon_-^2}\right),\\ \label{eq:anbnLin2}
   b_2&=\frac{1}{2} \sqrt{(1+2\Delta)  \left(\frac{1}{\epsilon_-^2}+\frac{1}{\epsilon_+^2}\right)+\frac{(\epsilon _--\epsilon _+)^2 \left(\epsilon _-^4+2 \epsilon _-^3 \epsilon _++6 \epsilon _-^2 \epsilon _+^2+2 \epsilon _- \epsilon _+^3+\epsilon _+^4\right)}{\epsilon _-^2 \epsilon _+^2 \left(\epsilon _-^2+\epsilon _+^2\right)^2}}.
\end{align}

Using the first few Lanczos coefficients, one can find the early time behavior of the Krylov complexity using Eq.~\eqref{eq:earlyCt}.
In the present case, the early time behaviour reads
\begin{equation} \label{eq:earlyCtLin}
     C(t)=\frac{\Delta}{4}  \left(\frac{1}{\epsilon_-^2}+\frac{1}{\epsilon_+^2}\right) t^2+\frac{\Delta(\epsilon _--\epsilon _+)^2}{16\;\epsilon _-^2 \epsilon _+^2 \left(\epsilon _-^2+\epsilon _+^2\right)}t^4+O(t)^6.
\end{equation}
This relation will be useful in the next section to compare with the holographic result.

It is interesting to note that using the definition of moments in terms of the return amplitude and their relation to the Lanczos coefficients \cite{Caputa:2024sux,Aguilar-Gutierrez:2025kmw}, one can find the relations
\begin{align} \label{eq:Ea0}
    E=\langle H \rangle &= a_0 = \frac{\Delta}{2}   \left(\frac{1}{\epsilon_-}+\frac{1}{\epsilon_+}\right),\\
    \langle H^2 \rangle&=b_1^2+a_0^2, \quad \Rightarrow \quad \delta E=\sqrt{\langle H^2 \rangle-\langle H \rangle^2}=b_1.
\end{align}
The energy obtained from the Lanczos coefficient methods here matches exactly with Eq.~\eqref{eq:linEn}.
Also, the standard deviation of the energy of the state is 
\begin{align} 
 \label{eq:Edev}
     \frac{\delta E}{E}&=\frac{b_1}{a_0}=\frac{\sqrt{\frac{1}{\epsilon_-^2}+\frac{1}{\epsilon_+^2}}}{ \sqrt{\Delta}\left(\frac{1}{\epsilon_-}+\frac{1}{\epsilon_+}\right)}=\frac{\sqrt{1+\tilde{\epsilon}^4}}{ \sqrt{\Delta}\left(1+\tilde{\epsilon}^2\right)},
\end{align}
where we used the Eq.~\eqref{eq:Linetaepm} to simplify. Since the function $f(\tilde{\epsilon})=\sqrt{1+\tilde{\epsilon}^4}/\left(1+\tilde{\epsilon}^2\right)$ is bounded for all $\tilde{\epsilon}$ values, $\sqrt{1/2}\leq f(\tilde\epsilon)<1$, Eq.~\eqref{eq:Edev} shows that one has a very sharp state with low deviation for large $\Delta$ values. 

In the limit $\epsilon_+=\epsilon_-=\epsilon$, one finds the well-known result for the return amplitude \cite{Caputa:2021sib, Caputa:2023vyr}
\begin{equation}
S(t)=\left(1-\frac{it}{2\epsilon}\right)^{-2\Delta}.
\end{equation}
with the analytic Lanczos coefficients
\begin{eqnarray}
a_n&=&\frac{(n+\Delta)}{\epsilon},\nonumber\\
b_n&=&\frac{1}{{2\epsilon}}\sqrt{n(n+2\Delta-1)},\label{anbnLO}
\end{eqnarray}
and an analytic spread complexity
\begin{equation} \label{eq:CtExLin}
\mathcal{C}(t)=\frac{\Delta }{2\epsilon^2}t^2.
\end{equation}

Some observations about the Lanczos coefficients in Eqs.(\ref{eq:anbnLin1}-\ref{eq:anbnLin2}) follows. The form of these coefficients does not fall in the family of the Lanczos coefficients that are governed by the $sl(2,\mathbb{R})$ algebra and are labeled by the highest weight representation $\Delta$. In that case one has a general form for $a_n=\gamma(n+\Delta)$ and for  $b_n=\alpha\sqrt{n(n+2\Delta-1)}$ \cite{Caputa:2021sib}\footnote{The spread complexity in that case is given as ~\cite{Balasubramanian:2022tpr}: 
$\mathcal{C}(t)=\frac{2\Delta}{1-\frac{\gamma^2}{4\alpha^2}}\sinh^2\left(t\sqrt{\alpha^2-\frac{\gamma^2}{4}}\right)$.}. The deviation from this behaviour is caused by the difference between $\epsilon_+$ and $\epsilon_-$ values and for the case that they are equal one finds agreement with the coefficients governed by $sl(2,\mathbb{R})$ in Eq.~(\ref{anbnLO}-\ref{eq:CtExLin}). Another point is that the generic coefficients are symmetric under $\epsilon_+$ and $\epsilon_-$ exchange, as the return amplitude is so.

Eq.~\eqref{eq:CtExLin} shows a quadratic behaviour of spread complexity for the case of $\epsilon_+=\epsilon_-$. It is interesting to note that there is a deviation from the quadratic behaviour of the spread complexity in Eq.~\eqref{eq:earlyCtLin}, at least at early times, for $\epsilon_+\neq\epsilon_-$ and the deviation is of course proportional to the $|\epsilon_+-\epsilon_-|$.

In the general case at hand, the coefficients with larger $n$ index get involved very quickly, and we can not deduce the general analytic expressions for $a_n$ and $b_n$, hence, we resort to numerical calculation. We numerically find the Lanczos coefficients up to $n\approx 4000$ for a reasonable convergence of the related spread complexity. The derivation has many technical details to obtain the coefficients with reasonable consumption of computational resources that are described in Appendix~\ref{app:Lanc}. There, we explain how the derivation of Lanczos coefficients from moments of return amplitude is an ill-conditioned problem. The method to overcome is using the power spectrum of the return amplitude instead of its moments and discretising the measure of integration for a stable and faster calculation.

\begin{figure}
    \centering
    \includegraphics[width=0.49\linewidth]{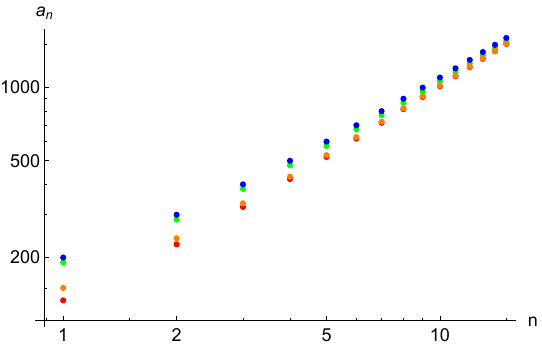}
    \includegraphics[width=0.49\linewidth]{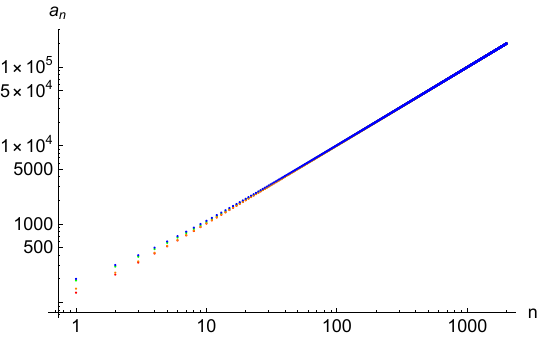} 
    \caption{Coefficients $a_n$, associated with the state on the plane, in the range $n=1-15$ (left) and $n=1-2000$ (right), for $\epsilon_+=1/100,\;\epsilon_-=1/100$ (Blue), $\epsilon_+=11/1000,\;\epsilon_-=1/100$ (Green), $\epsilon_+=2/100,\;\epsilon_-=1/100$ (Orange), $\epsilon_+=3/100,\;\epsilon_-=1/100$ (Red) and $\Delta=2$ in all cases. The case for $\epsilon_+=\epsilon_-$ matches with the exact relation in Eq.~\eqref{anbnLO}. Ln($a_n$) in plotted against Ln($n$). }
    \label{fig:anLin}
\end{figure}

\begin{figure}
    \centering
   \includegraphics[width=0.49\linewidth]{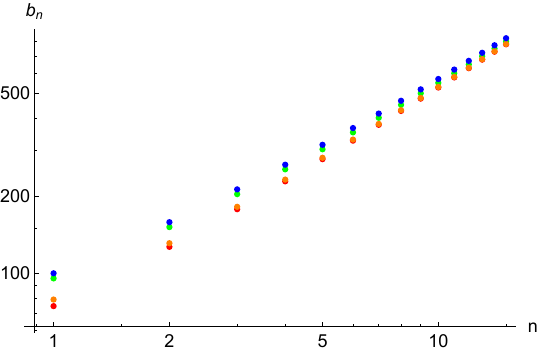}
    \includegraphics[width=0.49\linewidth]{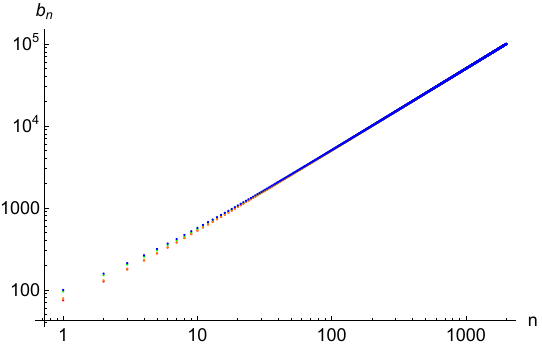}
    \caption{Coefficients $b_n$, associated with the state on the plane in the range $n=1-15$ (left) and $n=1-2000$ (right), for $\epsilon_+=1/100,\;\epsilon_-=1/100$ (Blue), $\epsilon_+=11/1000,\;\epsilon_-=1/100$ (Green), $\epsilon_+=2/100,\;\epsilon_-=1/100$ (Orange), $\epsilon_+=3/100,\;\epsilon_-=1/100$ (Red) and $\Delta=2$ in all cases. The case for $\epsilon_+=\epsilon_-$ matches with the exact relation in Eq.~\eqref{anbnLO}. Ln($b_n$) in plotted against Ln($n$).}
    \label{fig:bnLin}
\end{figure}

We provide plots for $a_n$ and $b_n$ coefficients associated with Eq.~\eqref{eq:StLin} in Figs.~\ref{fig:anLin}-\ref{fig:bnLin}. We observe that by keeping $\epsilon_-$ fixed and increasing $\epsilon_+$, the behaviour of both coefficients $a_n$ and $b_n$ changes considerably for small $n$, while for large $n$ the asymptote to the same value (See right panels of Figs.~\ref{fig:anLin}-\ref{fig:bnLin}). This is not accidental and relies on the analytical structure of the return amplitude in Eq.~\eqref{eq:StLin} in the complex $t$ plane \cite{Parker:2018yvk}. It is shown that for large $n$ the asymptotic behaviour of Lanczos coefficients is dictated by the position of the smallest pole of $S(t)$ on the imaginary $t$ axis, which happens at $it=2\epsilon_{min}=2\min\{\epsilon_+,\epsilon_-\}$. This is exactly the observed behaviour in the figures, as in all cases the Lanczos coefficients asymptote to the case with $\epsilon_+=\epsilon_-=\epsilon_{min}$. The values of $a_n$ and $b_n$ are bounded from above by $a_n$ and $b_n$ associated with $\epsilon_{min}$ applied to Eq.~\eqref{anbnLO}.

\begin{figure}
    \centering
    \includegraphics[width=0.7\linewidth]{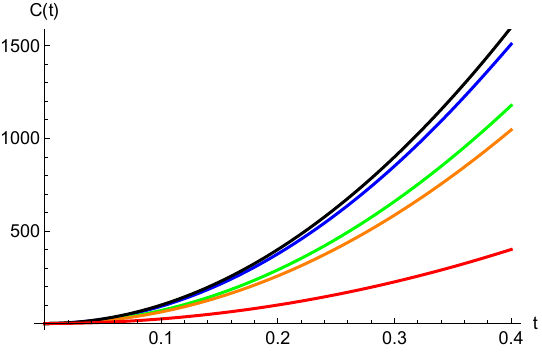}
    \caption{$C(t)$, for the state on the plane, for $\epsilon_+=1/100,\;\epsilon_-=1/100$ (Black),  $\epsilon_+=11/1000,\;\epsilon_-=1/100$ (Blue), $\epsilon_+=2/100,\;\epsilon_-=1/100$ (Green), $\epsilon_+=3/100,\;\epsilon_-=1/100$ (Orange), $\epsilon_+=2/100,\;\epsilon_-=2/100$ (Red) and $\Delta=2$ in all cases. The cases for $\epsilon_+=\epsilon_-$ match with the exact relation in Eq.~\eqref{eq:CtExLin}.}
    \label{fig:CtLin}
\end{figure}

The plots for the spread complexity for different values of $\epsilon_+$ and $=\epsilon_-$ are provided in Fig.~\ref{fig:CtLin} for chosen values of the parameters of the model. We see that by keeping $\epsilon_-$ fixed and increasing $\epsilon_+$, the value of spread complexity for all times is bounded from above by the one with  $\epsilon_-=\epsilon_+=\epsilon_{min}$. This guarantees that the growth of the complexity for the states under consideration will be at most like $C(t)\sim t^2$, and probably is in the same universality class as the complexity presented in Eq.~\eqref{eq:CtExLin} with $\epsilon_-=\epsilon_+$.

%%%%%%%%%
\subsection{CFT on the cylinder}  \label{sec:CFTCyl}
%%%%%%%%%
We now consider a 2d CFT on the cylinder. Let the
CFT live on a spatial circle of circumference $L$, with coordinates
$w=\tau+i\sigma$, $\bar{w}=\tau-i\sigma$, Euclidean time
$\tau\in(-\infty,+\infty)$ and $\sigma\sim\sigma+L$. Note that the choice of the complex coordinate convention is different compared to the previous section, as space is the imaginary part of $w, \bar w$. The map from the cylinder to the plane is given by the exponential maps
\begin{equation}
z(w)=e^{\frac{2\pi}{L}w}\,,\qquad \bar{z}(\bar{w})=e^{\frac{2\pi}{L}\bar{w}}\,.
\label{eq:CyltoLin}
\end{equation}
Similar to the previous section, we take as our state the vacuum
excited by a local primary and evolved in real time,
\begin{align}
\ket{\psi(t)}\equiv e^{-itH}\,e^{\frac{\epsilon_-+\epsilon_+}{2}H}e^{-\frac{\epsilon_+-\epsilon_-}{2}P}{\cal O}(0,0)\ket{0}\,,
\end{align}
with density matrix
\begin{equation} \label{eq:rhotCyl}
\rho(t)=\mathcal{N}e^{-iHt}{\cal O}(-\epsilon_-,-\epsilon_+)\ket{0}\bra{0}
{\cal O}^\dagger(\epsilon_-,\epsilon_+)e^{iHt}
\equiv \mathcal{N}\, {\cal O}(w_2,\bar{w}_2)\ket{0}\bra{0}{\cal O}^\dagger(w_1,\bar{w}_1)\,,
\end{equation}
where $\mathcal{N}$ enforces $\Tr\rho(t)=1$ and ${\cal O}$ is a spinless primary of
dimension $\Delta=2h$. As before, the Euclidean formalism treats $w$ and
$\bar{w}$ as independent variables, and the insertion points are
\begin{equation} \label{eq:cylcoor}
w_2=-\epsilon_-+it\,,\qquad \bar{w}_2=-\epsilon_++it\,,\qquad
w_1=\epsilon_-+it\,,\qquad \bar{w}_1=\epsilon_++it\,.
\end{equation}

% At $t=0$, the density matrix reads
% \begin{equation}\label{eq:rho0Cyl}
% \rho(0)= \mathcal{N}e^{-iHt}{\cal O}(-\epsilon_-,-\epsilon_+)\ket{0}\bra{0}
% {\cal O}^\dagger(\epsilon_-,\epsilon_+)e^{iHt}\,.
% \end{equation}
The one-point functions of the stress tensor components on a constant-time slice follow from the universal OPE. Writing $x_\pm=\sigma\pm t$ and evaluating
the holomorphic component at $w=i\sigma$ and the anti-holomorphic one at
$\bar{w}=-i\sigma$, one obtains \cite{Caputa:2026hxt}
\begin{eqnarray} \label{TpmGlobal}
\langle T_{\mp\mp}(x_\mp)\rangle\equiv \Tr\left(\rho(t)T_{\mp\mp}\right)&=&
\frac{\pi h}{2L^2}\frac{\sinh^2\left(\frac{2\pi\epsilon_\mp}{L}\right)}
{\sin^2\left(\frac{\pi(x_\mp+i\epsilon_\mp)}{L}\right)
\sin^2\left(\frac{\pi(x_\mp-i\epsilon_\mp)}{L}\right)}-\frac{c\pi}{12L^2}\,,
\end{eqnarray}
where the constant term is the usual Casimir contribution generated by the Schwarzian derivative of \eqref{eq:CyltoLin}. It was shown that this stress tensor matches the one obtained from the holographic background dual to this state~\cite{Caputa:2026hxt}. Integrating over the circle gives
the total energy and momentum,
% \begin{equation} \label{eq:CFTCylEJ}
% E=\frac{2\pi h}{L}\frac{\sinh\left(\frac{2\pi(\epsilon_++\epsilon_-)}{L}\right)}
% {\sinh\left(\frac{2\pi\epsilon_-}{L}\right)\sinh\left(\frac{2\pi\epsilon_+}{L}\right)}
% -\frac{c\pi}{6L}\,,\qquad
% J=\frac{2\pi h}{L}\frac{\sinh\left(\frac{2\pi(\epsilon_+-\epsilon_-)}{L}\right)}
% {\sinh\left(\frac{2\pi\epsilon_-}{L}\right)\sinh\left(\frac{2\pi\epsilon_+}{L}\right)}\,.
% \end{equation}
\begin{equation} \label{eq:CFTCylEJ}
E+\frac{c\pi}{6L}=\frac{2\pi h}{L}\left(\coth\frac{2\pi\epsilon_+}{L}+\coth\frac{2\pi\epsilon_-}{L}\right)\,,\qquad
J=\frac{2\pi h}{L}\left(\coth\frac{2\pi\epsilon_+}{L}-\coth\frac{2\pi\epsilon_-}{L}\right)\,,
\end{equation}
which shows that $\epsilon_+$ and $\epsilon_-$ control two chiralities independently, and in the decompactification limit $L\to\infty$ these reduce to $E\to h(\epsilon_+^{-1}+\epsilon_-^{-1})$ and $J\to h(\epsilon_-^{-1}-\epsilon_+^{-1})$, reproducing \eqref{eq:linEn} and
\eqref{eq:linMo}.

Similar to the previous section, in the limit $\epsilon_+=\epsilon_-=\epsilon$, the state describes an entangled
pair of excitations that is propagating to the right and the left of the origin at the speed of light, and the width of the smearing is set by the parameter $\epsilon$. 
In the generic case, two excitations will have different peak heights for left-moving and right-moving energy density pulses.

\textbf{Spread complexity:} Now, we focus on the introduction of the spread complexity of the associated state
\begin{equation}
\ket{\psi(0)}
=e^{\frac{\epsilon_-+\epsilon_+}{2}H}e^{-\frac{\epsilon_+-\epsilon_-}{2}P}{\cal O}(0,0)\ket{0}.
\end{equation}
Similar to the previous section, we start the computation from the return amplitude \cite{Caputa:2023vyr},
\begin{equation}
S(t)^*=\text{Tr}(\rho(0)e^{-iHt})=\frac{\langle \mathcal{O}^\dagger(w_3,\bar{w}_3)\mathcal{O}(w_4(t),\bar{w}_4(t))\rangle}{\langle\mathcal{O}^\dagger(w_3,\bar{w}_3) \mathcal{O}(w_4(0),\bar{w}_4(0))\rangle},
\end{equation}
where the insertion points are derived from Eqs.(\ref{eq:rhotCyl}) and (\ref{eq:cylcoor})
\begin{align}
w_3=-\epsilon_-,&\quad \bar{w}_3=-\epsilon_+,\nonumber\\
w_4(t)=\epsilon_- +it&,\quad \bar{w}_4(t)=\epsilon_+ +it.  
\end{align}
To proceed, one can either use the two-point correlator for a CFT on a cylinder with appropriate insertion points obtained from above or make use of Eq.~\eqref{eq:propLin} and perform the conformal transformation introduced in Eq.~\eqref{eq:CyltoLin} to obtain
\begin{equation}
S(t)=\left(\frac{\sinh\left(\frac{2\pi\epsilon_+}{L}\left(1-\frac{it}{2\epsilon_+}\right)\right)}{\sinh\left(\frac{2\pi\epsilon_+}{L}\right)}\right)^{-\Delta}\left(\frac{\sinh\left(\frac{2\pi\epsilon_-}{L}\left(1-\frac{it}{2\epsilon_-}\right)\right)}{\sinh\left(\frac{2\pi\epsilon_-}{L}\right)}\right)^{-\Delta}.\label{eq:CylSt}
\end{equation}
Note that the return amplitude is periodic with periodicity $L$.

Using this autocorrelation function, we can again derive the moments from Eq.~\eqref{eq:Stdef} that are used to find the Lanczos coefficients. Using the Toda hierarchy technique as described in Appendix~\ref{app:Lanc1}, we can find the analytic expressions for the Lanczos coefficients. Here we reproduce the first few coefficients keeping the convention $b_0=0$,
\begin{align}  \label{eq:anbnCyl1}
   a_0&=\frac{\pi}{L} \left({\mathfrak{e}_-}+{\mathfrak{e}_+}\right)\Delta,\\
   b_1&=\frac{\pi}{L} \sqrt{\Delta  \left(-2+{\mathfrak{e}_+}^2+{\mathfrak{e}_-}^2\right)},\\
   a_1&=\frac{\pi}{L}\left({\mathfrak{e}_-}+{\mathfrak{e}_+}\right)\left((\Delta+1)+\frac{({\mathfrak{e}_-}-{\mathfrak{e}_+})^2}{\left(-2+{\mathfrak{e}_+}^2+{\mathfrak{e}_-}^2\right)}\right),\\
   b_2&=\frac{\pi}{L} \Bigg((1+2\Delta)  \left(-2+{\mathfrak{e}_+}^2+{\mathfrak{e}_-}^2\right)+\nn\\
   &+\frac{({\mathfrak{e}_-}-{\mathfrak{e}_+})^2 \left({\mathfrak{e}_-}^4+2 {\mathfrak{e}_-}^3 {\mathfrak{e}_+}+6 {\mathfrak{e}_-}^2 \left({\mathfrak{e}_+}^2-1\right)+2 {\mathfrak{e}_-} {\mathfrak{e}_+}\left({\mathfrak{e}_+}^2-2\right)+{\mathfrak{e}_+}^4-6 {\mathfrak{e}_+}^2+4\right)}{ \left({\mathfrak{e}_-}^2+{\mathfrak{e}_+}^2-2\right)^2}\Bigg)^{1/2}  ,\label{eq:anbnCyl2}
\end{align}
where $\mathfrak{e}_+\equiv \coth{(\frac{2 \pi \epsilon_+}{L})}$ and $\mathfrak{e}_-\equiv \coth{(\frac{2 \pi \epsilon_-}{L})}$. In the $L\to\infty$ limit, one finds the results of the CFT on a plane from the previous section.

In this case, the early time behaviour of the Krylov complexity using Eq.~\eqref{eq:earlyCt} reads
\begin{align} \label{eq:earlyCtCyl}
     C(t)&=\frac{\pi^2\Delta}{L^2}  \left(-2+{\mathfrak{e}_+}^2+{\mathfrak{e}_-}^2\right)t^2\nn\\
    & +\frac{\pi ^4 \Delta }{3 L^4 \left({\mathfrak{e}_-}^2+{\mathfrak{e}_+}^2-2\right)} \left({\mathfrak{e}_-}^4 \left(3 {\mathfrak{e}_+}^2-4\right)-6 {\mathfrak{e}_-}^3 {\mathfrak{e}_+} \left({\mathfrak{e}_+}^2-1\right)+{\mathfrak{e}_-}^2 \left(3 {\mathfrak{e}_+}^4-8 {\mathfrak{e}_+}^2+7\right)\right.\nn\\
    &\left.+6 {\mathfrak{e}_-}{\mathfrak{e}_+} \left({\mathfrak{e}_+}^2-1\right)-4 {\mathfrak{e}_+}^4+7 {\mathfrak{e}_+}^2-4\right)t^4+O(t)^6.
\end{align}
This relation will be useful in Section~\ref{sec:Holography} to compare with the holographic result.

Again, in the limit $\epsilon_+=\epsilon_-=\epsilon$, one finds the well-known result for the return amplitude \cite{Caputa:2023vyr}
\begin{equation}
S(t)=\left(\frac{\sinh\left(\frac{2\pi\epsilon}{L}\left(1-\frac{it}{2\epsilon}\right)\right)}{\sinh\left(\frac{2\pi\epsilon}{L}\right)}\right)^{-2\Delta}
\end{equation}
with the analytic Lanczos coefficients
\begin{eqnarray}
a_n&=&\frac{2\pi\,(n+\Delta)}{L\tanh\left(\frac{2\pi\epsilon}{L}\right)},\nonumber\\
b_n&=&\frac{\pi}{L\sinh\left(\frac{2\pi\epsilon}{L}\right)}\sqrt{n(n+2\Delta-1)},\label{anbnCO}
\end{eqnarray}
and an analytic spread complexity
\begin{equation}\label{eq:CtExCy}
\mathcal{C}(t)=2\Delta\frac{\sin^2\left(\frac{\pi t}{L}\right)}{\sinh^2\left(\frac{2\pi \epsilon}{L}\right)}.
\end{equation}

Some comments about the Lanczos coefficients in Eqs.~(\ref{eq:anbnCyl1}-\ref{eq:anbnCyl2}) are as follows. Similar to the Eqs.~(\ref{eq:anbnLin1}-\ref{eq:anbnLin2}) in the previous section, the form of Lanczos coefficients does not fall into the form of the family of the Lanczos coefficients that are governed by the $sl(2,\mathbb{R})$ algebra. Hence, the general form for $a_n=\gamma(n+\Delta)$ and for  $b_n=\alpha\sqrt{n(n+2\Delta-1)}$ does not apply to this case either. The deviation from this behaviour is caused by the difference between $\epsilon_+$ and $\epsilon_-$ values and for the case that they are equal, one finds agreement with the coefficients governed by $sl(2,\mathbb{R})$ in Eq.~(\ref{anbnCO}-\ref{eq:CtExCy}). The coefficients are symmetric under $\epsilon_+$ and $\epsilon_-$ exchange, as the return amplitude is symmetric too.

The coefficients with larger $n$ index get involved very quickly, and we do not have the general analytic expressions for $a_n$ and $b_n$. Hence, we numerically find the Lanczos coefficients up to $n\approx 3000$ in order to produce the related spread complexity with a reasonable convergence in the whole period of motion. The derivation is similar to the previous section, and small differences due to the periodicity of the return amplitude are described in Appendix~\ref{app:Lanc}. %xyxy

\begin{figure}
    \centering
    \includegraphics[width=0.49\linewidth]{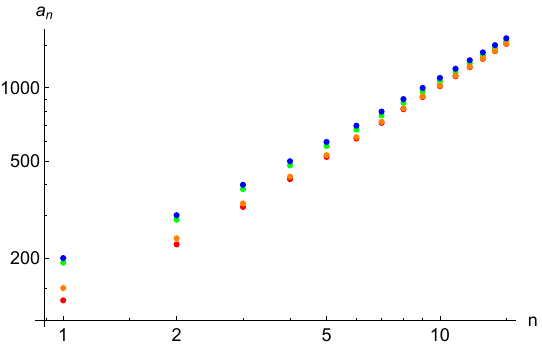}
    \includegraphics[width=0.49\linewidth]{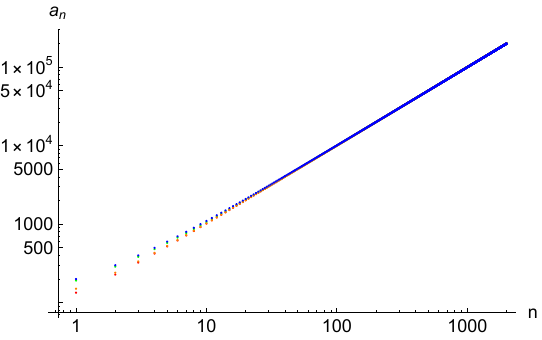} 
    \caption{Coefficients $a_n$, associated with the state on the cylinder in the range $n=1-15$ (left) and $n=1-2000$ (right), for $\epsilon_+=1/100,\;\epsilon_-=1/100$ (Blue), $\epsilon_+=11/1000,\;\epsilon_-=1/100$ (Green), $\epsilon_+=2/100,\;\epsilon_-=1/100$ (Orange), $\epsilon_+=3/100,\;\epsilon_-=1/100$ (Red) and $\Delta=2,\;L=1$ in all cases. The case for $\epsilon_+=\epsilon_-$ matches with the exact relation in Eq.~\eqref{anbnCO}. Ln($a_n$) in plotted against Ln($n$). }
    \label{fig:anCyl}
\end{figure}

\begin{figure}
    \centering
   \includegraphics[width=0.49\linewidth]{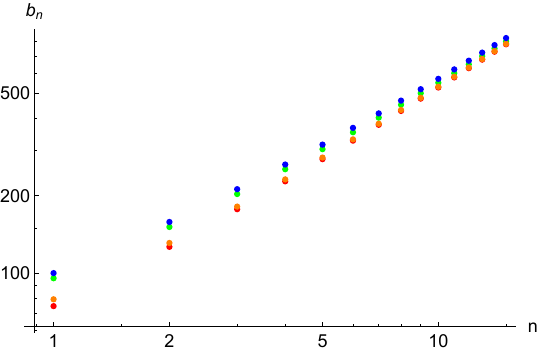}
    \includegraphics[width=0.49\linewidth]{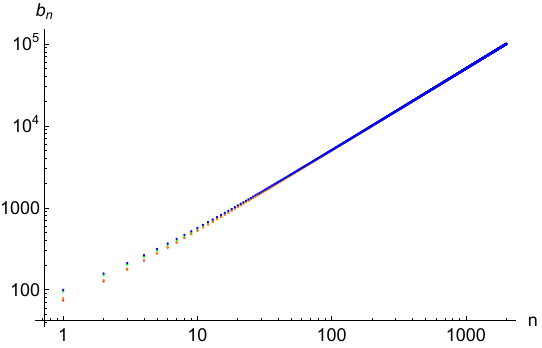}
    \caption{ Coefficients $b_n$, associated with the state on the cylinder in the range $n=1-15$ (left) and $n=1-2000$ (right), for $\epsilon_+=1/100,\;\epsilon_-=1/100$ (Blue), $\epsilon_+=11/1000,\;\epsilon_-=1/100$ (Green), $\epsilon_+=2/100,\;\epsilon_-=1/100$ (Orange), $\epsilon_+=3/100,\;\epsilon_-=1/100$ (Red) and $\Delta=2,\;L=1$ in all cases. The case for $\epsilon_+=\epsilon_-$ matches with the exact relation in Eq.~\eqref{anbnCO}. Ln($b_n$) in plotted against Ln($n$).}
    \label{fig:bnCyl}
\end{figure}

Now we provide plots for $a_n$ and $b_n$ associated with Eq.~\eqref{eq:CylSt} in Figs.~\ref{fig:anCyl}-\ref{fig:bnCyl}. Similar to the previous section, we observe that by keeping $\epsilon_-$ fixed and increasing $\epsilon_+$, the behaviour of both coefficients $a_n$ and $b_n$ changes considerably for small $n$, while for large $n$ they asymptote to the same value (See right panels of Figs.~\ref{fig:anCyl}-\ref{fig:bnCyl}). We check the analytical structure of the return amplitude in Eq.~\eqref{eq:CylSt} in the complex $t$ plane. For large $n$, the asymptotic behaviour of Lanczos coefficients is dictated by the position of the smallest pole in the imaginary $t$ axis for $S(t)$ happening at $2\epsilon_{min}=2\min\{\epsilon_+,\epsilon_-\}$. Again, this is exactly the observed behaviour in the figure, as in all cases the Lanczos coefficients asymptote to the case when $\epsilon_+=\epsilon_-=\epsilon_{min}$. The values of $a_n$ and $b_n$ are bounded from above by $a_n$ and $b_n$ associated with $\epsilon_{min}$ applied to Eq.~\eqref{anbnCO}.

\begin{figure}
    \centering
    \includegraphics[width=0.7\linewidth]{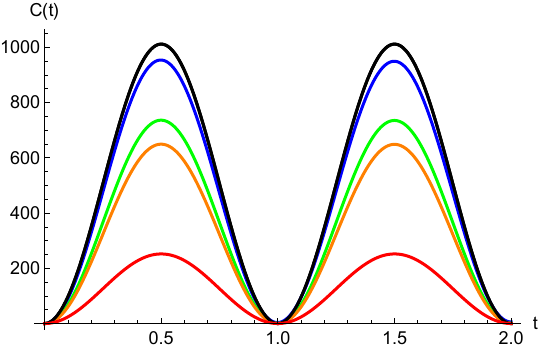}
    \caption{$C(t)$, for the state on the cylinder, for $\epsilon_+=1/100,\;\epsilon_-=1/100$ (Black), $\epsilon_+=11/1000,\;\epsilon_-=1/100$ (Blue),$\epsilon_+=2/100,\;\epsilon_-=1/100$ (Green),$\epsilon_+=3/100,\;\epsilon_-=1/100$ (Orange),$\epsilon_+=2/100,\;\epsilon_-=2/100$ (Red) and $\Delta=2,\;L=1$ in all cases. The cases for $\epsilon,_+=\epsilon_-$ match with the exact relation in Eq.~\eqref{eq:CtExCy}. }
    \label{fig:CtCyl}
\end{figure}

The plots for the spread complexity for different values of $\epsilon_+$ and $=\epsilon_-$ are provided in Fig.~\ref{fig:CtCyl} for chosen values of the parameters of the model. The spread complexity is periodic with the period determined by $L$. This is similar to the $\epsilon_-=\epsilon_+$ case with the analytic relation in Eq.~\eqref{eq:CtExCy}. The periodicity of the spread/Krylov complexity has been observed in different examples \cite{Caputa:2021sib,Fatemiabhari:2026goj}, and it is attributed to the finiteness of the dimension of the Hilbert space under consideration. The field theory is on a compact manifold with length $L$ that acts as an IR cut-off compared to the theory on a plane in the previous section, and the introduction of the smearing parameters $\epsilon_-$ and $\epsilon_+$ acts like a UV cut-off on the Hilbert space. Hence, one has a finite-dimensional subspace of the Hilbert space that bounds the length of the Krylov chain in Eq.~\eqref{eq:krylovBasis}. The state returns to the initial state $|\psi(0)\rangle$ after a certain time evolution, and the spread complexity vanishes at those moments.

We see that by keeping $\epsilon_-$ fixed and increasing $\epsilon_+$, the value of spread complexity for all times is bounded from above by the one associated with  $\epsilon_-=\epsilon_+$. This guaranties that the behaviour of the complexity for the states under consideration will be bounded from above by $\mathcal{C}(t)=2\Delta{\sin^2\left(\frac{\pi t}{L}\right)}/{\sinh^2\left(\frac{2\pi \epsilon_{min}}{L}\right)}$, and probably is in the same universality class as the complexity presented in Eq.~\eqref{eq:CtExCy} with $\epsilon_-=\epsilon_+$.


%%%%%%%%
\subsection{CFT at finite temperature}  \label{sec:CFTTem}
%%%%%%%%
For the last case to be studied, we consider a CFT on the thermal cylinder at inverse temperature $\beta$.  The cylinder is parametrised by periodic Euclidean time $\tau\sim\tau+\beta$ and non-compact spatial coordinate $x$. The complex coordinates can be defined as $(w,\bar{w})=(x+i\tau,\,x-i\tau)$ and the exponential maps to the plane is
\begin{equation}
z(w)=e^{\frac{2\pi}{\beta}w}\,,\qquad \bar{z}(\bar{w})=e^{\frac{2\pi}{\beta}\bar{w}}\,.
\label{eq:ThetoPlane}
\end{equation}
In place of the vacuum, we now excite the thermal density matrix $\rho_\beta=e^{-\beta H}$ by a local primary, again with independent chiral and anti-chiral regulators \cite{David:2026owc, Caputa:2026hxt},
\begin{equation} \label{eq:rhotThe}
\rho(t)=\mathcal{N}\,e^{-iHt}{\cal O}(-i\epsilon_-,i\epsilon_+)\,\rho_\beta\,
{\cal O}^\dagger(i\epsilon_-,-i\epsilon_+)\,e^{iHt}\,, 
\end{equation}
where $\mathcal{N}$ again enforces $\Tr\rho(t)=1$. Consistency of the Euclidean ordering requires
$0<\epsilon_\pm<\beta/2$, so that the two insertions lie within a single thermal period. For $\epsilon_+=\epsilon_-=\epsilon$, these states were studied as local operator quenches at finite temperature in \cite{Caputa:2014eta}.

Applying the universal OPE together with \eqref{eq:ThetoPlane}, the
stress-tensor one-point functions in light-cone coordinates
$x_\pm=x\pm t$ read
\begin{equation}
\langle T_{\pm\pm}(x_\pm)\rangle=\Tr\left(\rho(t)T_{\pm\pm}\right)
=\frac{\pi h}{2\beta^2}
\frac{\sin^2\left(\frac{2\pi\epsilon_\pm}{\beta}\right)}
{\sinh^2\left(\frac{\pi(x_\pm+i\epsilon_\pm)}{\beta}\right)
\sinh^2\left(\frac{\pi(x_\pm-i\epsilon_\pm)}{\beta}\right)}
+\frac{c\pi}{12\beta^2}\,,
\end{equation}
where the constant is the familiar thermal energy density of the unexcited state, generated by the Schwarzian derivative of \eqref{eq:ThetoPlane}.
As expected, each chirality is again controlled by its own regulator, and the expressions can be obtained from those of the previous subsection by the continuation $L\to i\beta$, which exchanges the trigonometric and hyperbolic functions and flips the sign of the central-charge term.

\textbf{Spread complexity:} Now we concentrate on the spread complexity calculation.
The return amplitude and other related quantities can be easily derived from the analytic continuation $L\to i \beta$ from the previous section or directly using the two-point function definition in Eq.~\eqref{eq:propLin} and the conformal transformation Eq.~\eqref{eq:ThetoPlane}.
Hence, for a CFT at finite temperature, we have the return amplitude associated with the density matrix $\rho(t)$ in Eq.~\eqref{eq:rhotThe} as
\begin{equation}
S(t)=\left(\frac{\sin\left(\frac{2\pi\epsilon_+}{\beta}\left(1-\frac{it}{2\epsilon_+}\right)\right)}{\sin\left(\frac{2\pi\epsilon_+}{\beta}\right)}\right)^{-\Delta}\left(\frac{\sin\left(\frac{2\pi\epsilon_-}{\beta}\left(1-\frac{it}{2\epsilon_-}\right)\right)}{\sin\left(\frac{2\pi\epsilon_-}{\beta}\right)}\right)^{-\Delta}.\label{eq:StThe}
\end{equation}

Using this autocorrelation function, we derive the moments that are used to find the Lanczos coefficients. Here again, we provide the first few terms with the convention $b_0=0$,
\begin{align} \label{eq:anbnThe1}
   a_0&=\frac{\pi}{\beta} \left({\bar {\mathfrak{e}}_-}+{\bar {\mathfrak{e}}_+}\right)\Delta,\\
   b_1&=\frac{\pi}{\beta} \sqrt{\Delta  \left(2+{\bar {\mathfrak{e}}_+}^2+{\bar {\mathfrak{e}}_-}^2\right)},\\
   a_1&=\frac{\pi}{\beta}\left({\bar {\mathfrak{e}}_-}+{\bar {\mathfrak{e}}_+}\right)\left((\Delta+1)+\frac{({\bar {\mathfrak{e}}_-}-{\bar {\mathfrak{e}}_+})^2}{\left(2+{\bar {\mathfrak{e}}_+}^2+{\bar {\mathfrak{e}}_-}^2\right)}\right),\\
   b_2&=\frac{\pi}{\beta} \Bigg((1+2\Delta)  \left(2+{\bar {\mathfrak{e}}_+}^2+{\bar {\mathfrak{e}}_-}^2\right)+\nn\\
   &+\frac{({\bar {\mathfrak{e}}_-}-{\bar {\mathfrak{e}}_+})^2 \left({\bar {\mathfrak{e}}_-}^4+2 {\bar {\mathfrak{e}}_-}^3 {\bar {\mathfrak{e}}_+}+6 {\bar {\mathfrak{e}}_-}^2 \left({\bar {\mathfrak{e}}_+}^2+1\right)+2 {\bar {\mathfrak{e}}_-} {\bar {\mathfrak{e}}_+}\left({\bar {\mathfrak{e}}_+}^2+2\right)+{\bar {\mathfrak{e}}_+}^4+6 {\bar {\mathfrak{e}}_+}^2+4\right)}{ \left({\bar {\mathfrak{e}}_-}^2+{\bar {\mathfrak{e}}_+}^2+2\right)^2}\Bigg)^{1/2} , \label{eq:anbnThe2}
\end{align}
where ${\bar {\mathfrak{e}}_+}\equiv \cot{(\frac{2 \pi \epsilon_+}{\beta})}$ and ${\bar {\mathfrak{e}}_-}\equiv \cot{(\frac{2 \pi \epsilon_-}{\beta})}$. 
In this case, the early time behaviour of the Krylov complexity using Eq.~\eqref{eq:earlyCt} reads
\begin{align}
     C(t)&=\frac{\pi^2\Delta}{L^2}  \left(2+{\bar {\mathfrak{e}}_+}^2+{\bar {\mathfrak{e}}_-}^2\right)t^2\nn\\
    & +\frac{\pi ^4 \Delta }{3 L^4 \left({\bar {\mathfrak{e}}_-}^2+{\bar {\mathfrak{e}}_+}^2+2\right)} \left({\bar {\mathfrak{e}}_-}^4 \left(3 {\bar {\mathfrak{e}}_+}^2+4\right)+6 {\bar {\mathfrak{e}}_-}^3 {\bar {\mathfrak{e}}_+} \left({\bar {\mathfrak{e}}_+}^2+1\right)+{\bar {\mathfrak{e}}_-}^2 \left(3 {\bar {\mathfrak{e}}_+}^4+8 {\bar {\mathfrak{e}}_+}^2+7\right)\right.\nn\\
    &\left.+6 {\bar {\mathfrak{e}}_-}{\bar {\mathfrak{e}}_+} \left({\bar {\mathfrak{e}}_+}^2+1\right)+4 {\bar {\mathfrak{e}}_+}^4+7 {\bar {\mathfrak{e}}_+}^2+4\right)t^4+O(t)^6.
\end{align}

In the limit $\epsilon_+=\epsilon_-=\epsilon$, one finds the relation for the return amplitude 
\begin{equation}
S(t)=\left(\frac{\sin\left(\frac{2\pi\epsilon}{\beta}\left(1-\frac{it}{2\epsilon}\right)\right)}{\sin\left(\frac{2\pi\epsilon}{\beta}\right)}\right)^{-2\Delta}
\end{equation}
with the analytic Lanczos coefficients and complexity as
\begin{eqnarray}
a_n&=&\frac{2\pi\,(n+\Delta)}{\beta\tan\left(\frac{2\pi\epsilon}{\beta}\right)},\nonumber\\
b_n&=&\frac{\pi}{\beta\sin\left(\frac{2\pi\epsilon}{\beta}\right)}\sqrt{n(n+2\Delta-1)},\label{anbnTO}
\end{eqnarray}
\begin{equation}\label{eq:CtExTe}
\mathcal{C}(t)=2\Delta\frac{\sinh^2\left(\frac{\pi t}{\beta}\right)}{\sin^2\left(\frac{2\pi \epsilon}{\beta}\right)}.    
\end{equation}

The Lanczos coefficients in Eqs.~(\ref{eq:anbnThe1}-\ref{eq:anbnThe2}) are qualitatively similar to the ones in the previous section. Again, the form of Lanczos coefficients does not fall in the family of the Lanczos coefficients that are governed by the $sl(2,\mathbb{R})$. The deviation from this behaviour is caused by the difference between $\epsilon_+$ and $\epsilon_-$ values, and for the case that they are equal, one finds agreement with the coefficients governed by $sl(2,\mathbb{R})$ in Eqs.~(\ref{anbnTO}-\ref{eq:CtExTe}).

Similar to the previous sections, we numerically find the Lanczos coefficients up to $n= 4000$ in order to produce the related spread complexity with a good convergence in the range of evolution. 

\begin{figure}
    \centering
    \includegraphics[width=0.49\linewidth]{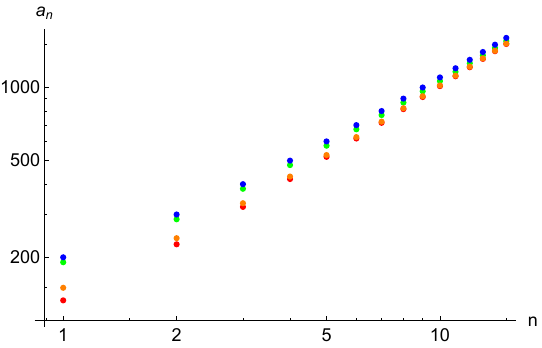}
    \includegraphics[width=0.49\linewidth]{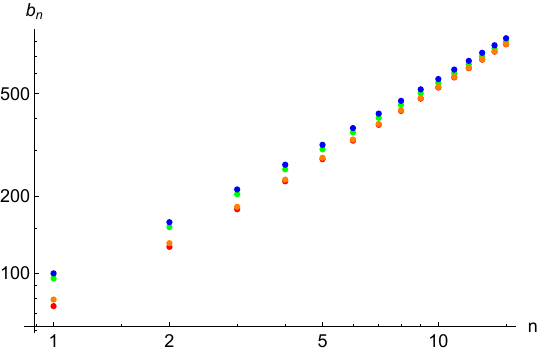} 
    \caption{ Coefficients $a_n$ (left) and $b_n$ (right), for the state in CFT at a finite temperature in the range $n=1-15$, for $\epsilon_+=1/100,\;\epsilon_-=1/100$ (Blue), $\epsilon_+=11/1000,\;\epsilon_-=1/100$ (Green), $\epsilon_+=2/100,\;\epsilon_-=1/100$ (Orange), $\epsilon_+=3/100,\;\epsilon_-=1/100$ (Red) and $\Delta=2,\;\beta=1$ in all cases. The case for $\epsilon_+=\epsilon_-$ matches with the exact relation in Eq.~\eqref{anbnTO}. Ln($a_n,b_n$) in plotted against Ln($n$). }
    \label{fig:anbnThe}
\end{figure}

Now we provide plots for $a_n$ and $b_n$ associated with Eq.~\eqref{eq:Stdef} in Fig.~\ref{fig:anbnThe}. Similar to the previous section, we observe that by keeping $\epsilon_-$ fixed and increasing $\epsilon_+$, the behaviour of both coefficients $a_n$ and $b_n$ is different for small $n$, while for large $n$ they asymptote to the same value. For large $n$, the asymptotic behaviour of $a_n$ and $b_n$ is dictated by the position of the smallest pole in the imaginary $t$ axis for $S(t)$ happening at $it=2\epsilon_{min}=2\min\{\epsilon_+,\epsilon_-\}$. The values of $a_n$ and $b_n$ are still bounded from above by $a_n$ and $b_n$ associated with $\epsilon_{min}$ applied to Eq.~\eqref{anbnTO}.

\begin{figure}
    \centering
    \includegraphics[width=0.7\linewidth]{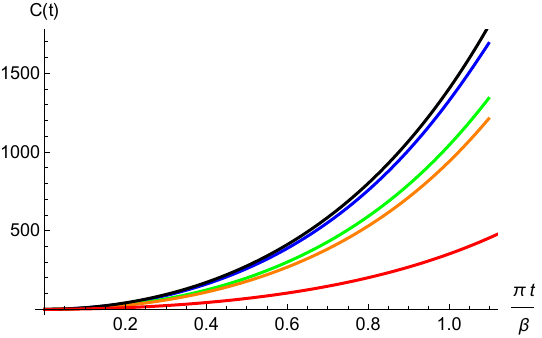}
    \caption{$C(t)$, associated with the state at finite temperature, for $\epsilon_+=1/100,\;\epsilon_-=1/100$ (Black), $\epsilon_+=11/1000,\;\epsilon_-=1/100$ (Blue),$\epsilon_+=2/100,\;\epsilon_-=1/100$ (Green),$\epsilon_+=3/100,\;\epsilon_-=1/100$ (Orange),$\epsilon_+=2/100,\;\epsilon_-=2/100$ (Red) and $\Delta=2,\;\beta=1$ in all cases. The cases for $\epsilon,_+=\epsilon_-$ match with the exact relation in Eq.~\eqref{eq:CtExTe}.}
    \label{fig:Ctbeta}
\end{figure}

The plots for the Krylov complexity for different values of $\epsilon_+$ and $=\epsilon_-$ are provided in Fig.~\ref{fig:Ctbeta} for chosen values of the parameters of the model. The spread complexity is growing exponentially, as is expected in finite temperature systems. This is similar to the $\epsilon_-=\epsilon_+$ case with the analytic relation in Eq.~\eqref{eq:CtExTe}. In this case, contrary to the CFT on a cylinder, the dimension of the relevant part of the Hilbert space under consideration is infinite.

We see that by keeping $\epsilon_-$ fixed and increasing $\epsilon_+$, the value of spread complexity for all times is bounded from above by the one associated with  $\epsilon_-=\epsilon_+$. This guarantees that the behaviour of the complexity for the states under consideration will be bounded from above by $\mathcal{C}(t)=2\Delta{\sinh^2\left(\frac{\pi t}{L}\right)}/{\sin^2\left(\frac{2\pi \epsilon_{min}}{L}\right)}$.

%%%%%
\section{The holographic approach} \label{sec:Holography}
%%%%%
In this section, we review the proposals for holographic duals to the spread complexity of states in field theory, in the context of the momentum of an infalling particle in the bulk. In the \cite{Caputa:2024sux}, the rate of change of the spread complexity of local quenches was matched with the proper momentum of a falling particle along its geodesic on the bulk side. This idea was generilsed as a conjecture in \cite{Chatzis:2026ekd,Chatzis:2026oou} to include cases where there is a conserved charge or momentum associated with the state in the field theory.
In this section, we briefly review the holographic setup related to the CFT states studied in the previous section based on \cite{Caputa:2026hxt}. Then we compare the rate of change of the spread complexity of these states with the proposed dual quantities. Appendix~\ref{app:gravity} provides a more detailed review of how a falling particle and its back reaction on the geometry in the bulk side are related to the dual state in the CFT side.

%%%%%%%
\subsection{Proper momentum}
In \cite{Caputa:2024sux}, the Krylov (or spread) complexity was computed for states excited by a primary operator $\mathcal{O}$ of dimension $\Delta$,
regulated by $e^{-\epsilon H}$ and thus carrying energy $E=\Delta/\epsilon$,
in a generic $\mathrm{CFT}_2$. This state corresponds to the case $\epsilon_+=\epsilon_-=\epsilon$ in Section~\ref{sec:CFTKry}. Similar to this study, three cases were treated: the CFT on the
cylinder at finite spatial size $L$; the CFT on the
plane with an infinite spatial line, and the CFT
at finite temperature $\beta=1/T$. The corresponding growth rates are
\begin{eqnarray}
\dot C_L(t)&=&\frac{\Delta}{\epsilon^2}\frac{L}{2\pi}\sin \frac{2\pi t}{L}\,,\label{KrylovL}\\
\dot C_\infty(t)&=&\frac{\Delta}{\epsilon^2}\, t\,,\label{Krylovinf}\\
\dot C_T(t)&=&\frac{\Delta}{\epsilon^2}\frac{\beta}{2\pi}\sinh \frac{2\pi t}{\beta}\,,\label{KrylovT}
\end{eqnarray}
where \eqref{Krylovinf} is recovered from \eqref{KrylovL} as $L\to\infty$. These results are obtainable from Eq.~\eqref{eq:CtExCy}, Eq.~\eqref{eq:CtExLin},  and Eq.~\eqref{eq:CtExTe} respectively.

These results were matched by bulk computations in the global $AdS_3$, in Poincar\'{e} $AdS_3$, and for the
$AdS_3$ black hole
\begin{eqnarray} \label{eq:glomet}
ds_{\rm 3,gl.}^2&=& R^2d\rho^2 +\frac{4\pi^2R^2}{L^2}\left(-\cosh^2 \rho \,dt^2+\sinh^2\rho\, d\phi^2\right)\,,\\\label{eq:poinmet}
ds_{\rm 3,P.}^2&=& R^2d\rho^2+R^2e^{2\rho}\left(-dt^2+dx^2\right)\,,\\\
ds_{\rm 3,BH}^2&=& R^2d\rho^2+\frac{4\pi^2R^2}{\beta^2}\left(-\sinh^2\rho \, dt^2+\cosh^2\rho\,dx^2\right)\,,
\end{eqnarray}
where $\rho$ is chosen so that $ds^2=d\rho^2$ at fixed other spatial boundary coordinates,
i.e.\ $\rho$ is proper radial distance. The canonical momentum $P_\rho$ conjugate to this coordinate was termed the \emph{proper momentum}. Releasing a
massive particle from rest deep in the UV, at $z=e^{-\rho}=\epsilon$, and following its geodesic, one finds in all cases
\begin{equation} \label{eq:Cdot}
\dot C(t)=-\frac{P_\rho(t)}{\epsilon}\,.
\end{equation}

It was proposed in \cite{Fatemiabhari:2025cyy} that \eqref{eq:Cdot} continues to hold away from $AdS_3$, provided one correctly identifies the proper distance coordinate $y$, defined by $ds^2=dy^2$ along the trajectory of a massive particle with $P_{y}$ the
canonical momentum conjugate to it.

Concretely, imagine a probe particle that is moving in a two-dimensional subspace of the original background parametrised by $(r,\tilde x)$ coordinates. By  parametrising the timelike geodesic with $\left(r(t),\tilde x(t)\right)$, the line element restricted to the trajectory takes the form
\begin{equation}
ds^2=A(t)\,dr^2+B(t)\,d{\tilde x}^2\equiv dy^2\,,  
\end{equation}
where $A(t)$ and $B(t)$ are the metric functions evaluated along the geodesic, $A(t)=A\left(r(t),{\tilde x}(t)\right)$ and likewise for $B$. Dividing by $dt^2$ provides the rate of change of proper distance,
\begin{equation}
\frac{dy}{dt}=\sqrt{A\,\dot r^2+B\,\dot{\tilde x}^2}\,,
\end{equation}
so that
\begin{equation}
\frac{dr}{dy}=\frac{\dot r}{\sqrt{A\,\dot r^2+B\,\dot{\tilde x}^2}}\,,
\qquad
\frac{d{\tilde x}}{dy}=\frac{\dot{\tilde x}}{\sqrt{A\,\dot r^2+B\,\dot{\tilde x}^2}}\,.
\end{equation}
The proper momentum then follows by the chain rule,
\begin{equation}
P_{y}=P_r\frac{dr}{dy}+P_{\tilde x}\frac{d{\tilde x}}{dy}
=\frac{P_r\,\dot r+P_{\tilde x}\,\dot{\tilde x}}{\sqrt{A\,\dot r^2+B\,\dot{\tilde x}^2}}\,,
\end{equation}
which reduces to $P_{y}=P_{\rho}$ in the $AdS_3$ cases where the geodesic was purely radial. 

Immediately one can observe that if at $t=0$, $\dot r(0)\neq0$ or $\dot {\tilde x}(0)\neq0$, then $P_{y}$ at $t=0$ will have a non zero value.
Then, following the proposal 
\begin{equation}
    P_{y}(t)=P_0+O(t)\quad \rightarrow \quad C(t)\propto \int dt\;P_{y}(t)=P_0 t + O(t)^2.
\end{equation}
This will be in contradiction with the general result presented in Eq.~\eqref{eq:earlyCt} that the spread/ Krylov complexity at early times should be an even function of time. Hence, this proposal is contradictory with the field theory by definition. Since in all cases we are going to consider later there is a non-zero initial momentum for the probe particle, we move on to review an amended proposal. 

\subsection{The Routhian method}
In this section, we follow \cite{Chatzis:2026ekd,Chatzis:2026oou}. Consider a probe governed by a Lagrangian
$\mathcal{L}(r,\dot r,{\tilde x},\dot{\tilde x})$ in which ${\tilde x}$ is cyclic, so that the
conjugate charge
\begin{equation}
 J=\frac{\partial\mathcal{L}}{\partial\dot{\tilde x}}
\end{equation}
is conserved and may be held fixed.

The proposed resolution of the problem raised above is to carry out the fixed-charge reduction \emph{before} calculating the bulk observable dual to the complexity. Having the
constraint solved for $\dot{\tilde x}=\dot{\tilde x}(r,\dot r;J)$, one performs a
Legendre transform to the Routhian,
\begin{equation}\label{eq:intro_routhian}
 \mathcal{R}(r,\dot r;J)=\left[J\dot{\tilde x}-\mathcal{L}(r,\dot r,{\tilde x},\dot{\tilde x})\right]
 \big|_{\dot{\tilde x}=\dot{\tilde x}(r,\dot r;J)}\,.
\end{equation}
The Routhian generates the same geodesic, but some charge-dependent effective mass, or effective potential, takes the place of the cyclic motion-related coordinate. Then, one can extract the proper coordinate and its conjugate momentum from $\mathcal{R}$. The momentum obtained in this way will be purely radial and vanishes at the initial point in agreement with the universal early-time growth \eqref{eq:earlyCt}. In the next section, we will apply this proposal to the probe particles dual to the states studied in Section~\ref{sec:CFTKry}.

%%%%%%%%%%%%
\subsubsection{Poincare AdS}
%%%%%%%%%%%%
Consider a probe particle in AdS$_3$ parametrized by Poincare coordinates that at $t=0$ has an initial momentum in the spatial direction $x$ but not in the radial direction in the background of Eq.~\eqref{eq:poinmet}. In Ref.~\cite{Caputa:2026hxt}, it was shown that the back-reacted geometry obtained from this particle reproduces the correct holographic energy-momentum tensor associated with the state introduced in Section~\ref{sec:CFTLin}. More details about the geometry are discussed in Appendix~\ref{app:gravity}. We apply the proposal of proper momentum, discussed in the previous sections, to this particle and compare it with the results in Section~\ref{sec:CFTLin}.

We start by considering the  metric in the $z$ radial coordinate
\begin{equation}
ds^2=R^2\frac{-dt^2+dx^2+dz^2}{z^2}\,.\label{eq:metPo}
\end{equation}

We want to obtain the geodesic $(x(t),z(t))$ of a massive particle falling in this geometry with non-zero initial velocity in the $x$ direction. The action is
\begin{equation}
S_m=-mR\int dt\frac{\sqrt{1-\dot x(t)^2-\dot z(t)^2}}{z(t)}\equiv\int dt\mathcal{L}\,.
\end{equation}
For the initial conditions, we have
\begin{equation}
x(t=0)=0,\qquad \dot x(0)=v,\qquad  z(t=0)=\epsilon,\qquad \dot z(0)=0\,.\label{eq:IVPoincare}
\end{equation}
The solution reads
\begin{equation}
x(t)=vt\,,\qquad z(t)=\sqrt{\left(1-v^2\right)t^2+\epsilon^2}\,.\label{eq:geoPoincare}
\end{equation}
This gives the particle's trajectory in terms of its charges and the cut-off.
From $z(0)=\epsilon$ and $\dot x(0)=v$, one can fix the relation to the boost parameters introduced in Appendix~\ref{app:gravity}
\begin{equation}
Re^{\eta_1}=\epsilon,\qquad \tanh(\eta_2)=v\,.\label{eq:boostPoin}
\end{equation}
The energy and momentum of the particle are
\begin{equation} \label{eq:LinEP}
E=\frac{mR}{\epsilon\sqrt{1-v^2}}=me^{-\eta_1}\cosh(\eta_2)\,,\qquad P=\frac{mRv}{\epsilon\sqrt{1-v^2}}=me^{-\eta_1}\sinh(\eta_2)\,,
\end{equation}
Introducing the relations
\begin{equation}
\Delta=2h=mR\,,\qquad \epsilon_\pm=R\,e^{\eta_1\pm\eta_2}\,,
\label{eqn:LinCFTGrav}
\end{equation}
one can see that the energy and momentum of the particle match with the energy and momentum attributed to the excitation in CFT on the plane of Eq.~\eqref{eq:linEn}-\eqref{eq:linMo}. This completes the map between boost parameters, initial value data, and the CFT state smearing variables.

Now we calculate the proper momentum of the particle using the Routhian method.

For our Lagrangian, one has two conserved quantities
\begin{equation}
    {\cal P} = \frac{\partial {\cal L}}{\partial \dot{x}}\quad \& \quad {\cal H}=\frac{\partial {\cal L}}{\partial \dot{z}}\dot{z}+{\cal P}\dot{x} - {\cal L},
\end{equation}
that on-shell they reduce to Eq.~\eqref{eq:LinEP}.
We perform a Legendre transform with respect to the coordinate $x$ in order to put the system in the sector of fixed momentum $\mathcal{P}$. The resulting Routhian gives the relevant effective action for the fixed conserved momentum dynamics. One uses (we set $R=1$ for simplicity)
\begin{equation}
    \dot{x}=\frac{{\cal P}z\sqrt{1- \dot{z}^2}}{  \sqrt{ m^2 + {\cal P}^2 z^2}},
\end{equation}
to eliminate the dependence on the $\dot{x}$ coordinate,
\begin{equation}
    {\cal L}(z,\dot z,\dot{x})\mapsto {\cal R}(z,\dot{z},{\cal P}) = {\cal P}\dot{x}(z,\dot{z},{\cal P}) - {\cal L}(z,\dot{z},{\cal P}).
\end{equation}
Hence, the Routhian is
\begin{equation}\label{eq:RuthianLin}
    {\cal R} = \frac{\sqrt{\left(\left(1-\dot z^2\right) \left(m^2+{\cal P}^2 z^2\right)\right)}}{z}.
\end{equation}
We can now calculate the proper momentum using the Routhian instead of the Lagrangian,
\begin{equation}
    {\cal R} = {\sqrt{\left(1/z^2\right) \left(m^2+{\cal P}^2 z^2\right)-\dot y^2}},\quad \dot{y}=\frac{\sqrt{m^2+{\cal P}^2 z^2}}{z}\dot{z},
\end{equation}
which gives 
\begin{equation}
    P_y \equiv - \frac{\partial {\cal R}_p}{\partial \dot{y}}=\frac{\dot{y}}{\sqrt{\left(1/z^2\right) \left(m^2+{\cal P}^2 z^2\right)-\dot{y}^2}}=\frac{\dot{z}}{\sqrt{1 - \dot{z}^2}}.
\end{equation}
Substituting the solution for $z(t)$ from Eq.~\eqref{eq:geoPoincare} one has
\begin{equation}
   P_y = \frac{t-t v^2}{\sqrt{t^2 \left(v^2-v^4\right)+\epsilon ^2}}.
\end{equation}

Now we can expand the expression for early time and integrate to obtain $C(t)=\Lambda\int P_y \;dt$, where $\Lambda$ is a proportionality constant that may help us to match with the CFT result in Eq.~\eqref{eq:earlyCtLin},
\begin{equation}
    \Lambda\int P_y \;dt=\Lambda\frac{t^2 \left(1-v^2\right)}{2 \epsilon }-\Lambda\frac{t^4 \left(v^2 \left(v^2-1\right)^2\right)}{8 \epsilon ^3}+O\left(t^6\right).
\end{equation}
By using the relations in Eqs.~\eqref{eq:boostPoin} and \eqref{eqn:LinCFTGrav} one gets
\begin{align} \label{eq:propC}
     \Lambda\int P_y \;dt=\frac{\Delta   \left(1/\epsilon_-^2+1/\epsilon_+^2\right)}{4 }t^2-\frac{\Delta   (\epsilon_--\epsilon_+)^2 \left(\epsilon_-^2+\epsilon_+^2\right)}{4 \epsilon_-^2 \epsilon_+^2 (\epsilon_-+\epsilon_+)^4}t^4+O(t)^6.
\end{align}
where $\Lambda=\frac{\Delta  (\epsilon_-+\epsilon_+)^2 \left(\epsilon_-^2+\epsilon_+^2\right)}{8 (\epsilon_- \epsilon_+)^{5/2}}$ is chosen to match the first term in the expansion with Eq.~\eqref{eq:earlyCtLin}. One can see that the higher-order term in Eq.~\eqref{eq:earlyCtLin}, rewritten here for convenience,
\begin{equation}
     C(t)=\frac{\Delta}{4}  \left(\frac{1}{\epsilon_-^2}+\frac{1}{\epsilon_+^2}\right) t^2+\frac{\Delta(\epsilon _--\epsilon _+)^2}{16\;\epsilon _-^2 \epsilon _+^2 \left(\epsilon _-^2+\epsilon _+^2\right)}t^4+O(t)^6. \nn
\end{equation}
does not match with the proposal in Eq.~\eqref{eq:propC}.

\subsubsection{Global AdS}
%%%%%%%%%%
For a probe particle in AdS$_3$ background of Eq.~\eqref{eq:glomet} that at $t=0$ has an initial momentum in the spatial direction $\phi$ but not in the radial direction, Ref.~\cite{Caputa:2026hxt}, shows that the back-reacted geometry obtained from this particle reproduces the correct holographic energy momentum tensor associated with the state introduced in Section~\ref{sec:CFTCyl}. More details about the geometry are provided in Appendix~\ref{app:gravity}. We apply the proposal of proper momentum to this particle and compare it with the results in Section~\ref{sec:CFTCyl}.
We repeat the metric here
\begin{equation}
ds^2=R^2\left[d\rho^2+\frac{4\pi^2}{L^2}\left(-\cosh^2(\rho)dt^2+\sinh^2(\rho)d\phi^2\right)\right]\,.
\end{equation}
We want to obtain the geodesic $(\phi(t),\rho(t))$ of a massive particle falling in this geometry with non-zero initial velocity in the $\phi$ direction, 
\begin{equation}
S_m=-mR\int dt\sqrt{\frac{4\pi^2}{L^2}(\cosh^2(\rho(t))-\sinh^2(\rho(t))\dot\phi(t)^2)-\dot\rho^2(t)}\,=\int dt {\cal L}. 
\end{equation}
For the initial conditions 
\begin{equation}
\phi(0)=0\,,\qquad \dot\phi(0)=\omega\,,\qquad \rho(0)=\rho_\Lambda\,,\qquad \dot\rho(0)=0\,,
\end{equation}
The solution reads \cite{Caputa:2026hxt},
\begin{equation}
\phi(t)=\frac{L}{2\pi}\tan^{-1}\left(\omega\tan\left(\frac{2\pi t}{L}\right){\tanh^2(\rho_\Lambda)}\right)\,,
\end{equation}
\begin{align} \label{eq:rhotglobal}
\rho(t)&=\tanh^{-1}\left(\cos\left(\frac{2\pi t}{L}\right)\tanh(\rho_\Lambda)\sqrt{1+\omega^2\tan^2\left(\frac{2\pi t}{L}\right)}\right).
\end{align}
This gives the particle's trajectory in terms of its charges and the UV cut-off. From $\dot\phi(0)=\omega\,, \rho(0)=\rho_\Lambda\,,$ one can fix the relation to the boost parameters introduced in Appendix~\ref{app:gravity},
\begin{align} \label{eq:Cylequi}
\rho_\Lambda\,&=\eta_1,\\
\omega&=\frac{\tanh(\eta_2)}{\tanh(\eta_1)}\,.    \label{eq:Cylequi2}
\end{align}
Using relations above, the energy and angular momentum of the particle read
\begin{equation}
E=\frac{2\pi m R}{L}\cosh(\eta_1)\cosh(\eta_2)\,,\qquad J=\frac{2\pi m R}{L}\sinh(\eta_1)\sinh(\eta_2)\,,\label{eq:globarEJ}
\end{equation}
that can be matched with Eq.~\eqref{eq:CFTCylEJ} upon identifications
\begin{align} 
\tanh\left(\frac{\pi\epsilon_-}{L}\right)=e^{-(\eta_1+\eta_2)}\,,\qquad\tanh\left(\frac{\pi\epsilon_+}{L}\right)=e^{-(\eta_1-\eta_2)}\,,\label{eq:globaleta}
\end{align}
and considering central charge and stress tensor normalisations.
We will make use of these parametrisation to compare the complexity proposal with the CFT result.

Now we focus on calculating the proper momentum of the particle using the Routhian method.
Again, for our Lagrangian, one has two conserved quantities
\begin{equation}
    {\cal J} = \frac{\partial {\cal L}}{\partial \dot{\phi}}\quad \& \quad {\cal H}=\frac{\partial {\cal L}}{\partial \dot{\rho}}\dot{\rho}+{\cal J}\dot{\phi} - {\cal L},
\end{equation}
that on-shell reduces to the angular momentum and energy of the particle provided in Eq.~\eqref{eq:globarEJ}.

We perform a Legendre transform with respect to the coordinate $\phi$. The resulting Routhian gives the relevant effective action for the fixed conserved angular momentum dynamics. Explicitly, in this case, one substitutes (we set $R=1, L=1$ for simplicity),
\begin{equation}
    \dot{\phi}=\frac{{\cal J} \cosh(\rho) \sqrt{4 \pi ^2 \cosh ^2(\rho)-\dot\rho^2}}{\pi  \sqrt{4 {\cal J}^2+16 \pi ^2 m^2 \sinh ^2(\rho)}},
\end{equation}
such that the Routhian does not depend on the coordinate $\dot{\phi}$,
\begin{equation}
    {\cal L}(\rho,\dot\rho,\dot{\phi})\mapsto {\cal R}(\rho,\dot{\rho},{\cal J}) = {\cal J}\dot{\phi}(\rho,\dot{\rho},{\cal J}) - {\cal L}(\rho,\dot{\rho},{\cal J}).
\end{equation}
We have
\begin{equation}\label{eq:RuthianGlobal}
    {\cal R} = \frac{\sqrt{\left(4 \pi ^2 \cosh ^2(\rho)-\dot \rho^2\right) \left({\cal J}^2 \cosh^2(\rho)+4 \pi ^2 m^2\right)}}{2 \pi }.
\end{equation}
The Routhian reads
\begin{equation}
    {\cal R} = \sqrt{\cosh ^2(\rho) \left({\cal J}^2 \cosh^2(\rho)+4 \pi ^2 m^2\right)-\dot y^2} ,\quad \dot{y}=\frac{\sqrt{ \left({\cal J}^2 \cosh^2(\rho)+4 \pi ^2 m^2\right)}}{2 \pi}\dot{\rho},
\end{equation}
which gives 
\begin{equation}
    P_y \equiv - \frac{\partial {\cal R}_p}{\partial \dot{y}}=\frac{\dot{y}}{\sqrt{\cosh ^2(\rho) \left({\cal J}^2 \cosh^2(\rho)+4 \pi ^2 m^2\right)-\dot y^2}}=\frac{\dot \rho}{\sqrt{4 \pi ^2 \cosh ^2(\rho)-\dot \rho^2}}.
\end{equation}
Substituting the solution for $\rho(t)$ from Eq.~\eqref{eq:rhotglobal} one has
\begin{equation}
   P_y = \frac{\left(\omega ^2-1\right) \tanh (\rho_\Lambda) \sin (2 \pi  t)}{\sqrt{-\tanh ^2(\rho_\Lambda) \left(\omega ^4 \tan ^2(2 \pi  t)+1\right)+\omega ^2 \tan ^2(2 \pi  t)+1}}.
\end{equation}

Now we can expand this expression for early time and integrate to obtain $C(t)=\Lambda\int P_y \;dt$, where $\Lambda$ is the proportionality constant that can help us to match with the CFT result.
\begin{align}
    \Lambda\int P_y \;dt&=\Lambda\pi  t^2 \left(\omega ^2-1\right) \sinh (\rho_\Lambda)\nn\\
    &+\Lambda\frac{1}{3} \pi ^3 t^4 \left(\omega ^2-1\right) \sinh (\rho_\Lambda) \cosh ^2(\rho_\Lambda) \left(\left(3 \omega ^4+1\right) \tanh ^2(\rho_\Lambda)-3 \omega ^2-1\right)+O\left(t^6\right)
\end{align}
By using the relations in Eqs.~\eqref{eq:Cylequi}-\eqref{eq:Cylequi2} one gets
\begin{align} \label{eq:Pycyl}
     \Lambda\int P_y \;dt&=\pi ^2 \Delta  (\cosh (2 \eta_1) \cosh (2 \eta_2)-1)t^2\nn\\
     &+\frac{1}{48} \pi ^4 \Delta  \text{sech}^2(\eta_2) \left(-4 \cosh (2 \eta_1) \cosh (4 \eta_2)-(13 \cosh (2 \eta_1)+14 \cosh (4 \eta_1)\right.\nn\\
     &\left.+3 (\cosh (6 \eta_1)+6)) \text{csch}^2(\eta_1) \cosh (2 \eta_2)-96 \cosh ^4(\eta_1) \coth ^2(\eta_1) \text{sech}^2(\eta_2)\right.\nn\\
     &\left.+4 \left(41 \cosh (2 \eta_1)+6 \cosh (4 \eta_1)+36 \text{csch}^2(\eta_1)+74\right)\right)t^4+O(t)^6
\end{align}
where 
\begin{equation}
\Lambda=-\frac{2 \pi  \Delta  \sinh (\eta_1) \cosh ^2(\eta_2) (\cosh (2 \eta_1) \cosh (2 \eta_2)-1)}{\cosh (2 \eta_1)-\cosh (2 \eta_2)},
\end{equation}
is chosen to match the first term in the expansion with the Eq.~\eqref{eq:earlyCtCyl}. But Eq.~\eqref{eq:earlyCtCyl} in this parameters will be
\begin{align}
    C(t)&=\pi ^2 \Delta  (\cosh (2 \eta_1) \cosh (2 \eta_2)-1)t^2\nn\\
     &\frac{\pi ^4 \Delta }{48 (\cosh (2 \eta_1) \cosh (2 \eta_2)-1)}\left(6 (2 \cosh (4 \eta_1)-1) \cosh ^2(\eta_1) \cosh (2 \eta_2)\right.\nn\\
     &\left.-2 (5 \cosh (4 \eta_1)+2) \cosh (4 \eta_2)+\cosh (2 \eta_1) (26 \cosh (2 \eta_2)+6 \cosh (4 \eta_2)-3)\right.\nn\\
     &\left.+6 \sinh ^2(\eta_1) \cosh (6 \eta_2)-4 \cosh (4 \eta_1)-3 \cosh (6 \eta_1)-14\right)t^4+O(t)^6
\end{align}
One can see that the higher-order term in this equation does not agree with the proposal in Eq.~\eqref{eq:Pycyl}.

The procedure to obtain results for the particle falling in BTZ geometry dual to the state studied in Section.~\ref{sec:CFTTem} is similar, and we do not reproduce them here.

\section{Conclusions and Outlook}\label{conclsect}
Using the universal two-point correlation functions in 2d CFTs, we have studied the return amplitude of locally excited states with independent chiral and anti-chiral regulators, $\epsilon_\pm$, in three different set-ups: CFT on a plane, CFT on a cylinder, and CFT at finite temperature. Using these exact return amplitudes, we managed to obtain a few Lanczos coefficients and the early time behaviour of the spread complexity of these states analytically.

Then we used spectral methods and discretisation of the measure of the power spectrum of our return amplitudes to find the $a_n$ and $b_n$ coefficients up to high orders in Krylov index, $n$, numerically, and use them to calculate the spread complexity of the excited states during a longer period of time evolution. This method helped us to improve convergence for the observable values under consideration using reasonable computational resources.

The coefficients $a_n$ and $b_n$ in all three cases depend on values of $\epsilon_+$ and $\epsilon_-$ non-trivially, but in the large $n$ limit they asymptote to the values of coefficients associated with the state with symmetric regulator $\epsilon_{min}=\min\{\epsilon_+,\epsilon_-\}$.  The coefficients do not fit into the form dictated by symmetry algebra $sl(2,\mathbb{R})$, hence their closed form was not known to us.

The spread complexity has different behaviours in different cases. For CFT on the plane and at finite temperature, the complexity for generic $\epsilon_+$ and $\epsilon_-$ grows unbounded, and specially at finite temperature, we observe the exponential growth. For CFT on a cylinder, the complexity is periodic with the same period as the $\epsilon_+=\epsilon_-$ case, and this can be attributed to the finiteness of the Hilbert space accessible to the state under consideration. In all cases, the values of the spread complexity are bounded from above by one of $\epsilon_+=\epsilon_-=\epsilon_{min}$.

The analytical early time behaviour of the complexity helps us to compare with the holographic proposals of the spread complexity. Intriguingly, we do not get agreement for the rate of change of complexity being proportional to the proper momentum of the falling particle, whether this momentum is obtained from the Lagrangian of the particle or the Routhian. This suggests that the holographic dual to this observable should be amended.

Ageev et al.~\cite{Ageev:2018nye} studied the complexity-equals-volume (CV) and complexity-equals-action (CA) prescriptions for a holographic local quench in a 2d CFT. They found that the subsystem volume complexity can decrease in certain regimes while the entanglement entropy and entanglement density continue to grow. They interpreted this as a signature of effective (Gell-Mann-Lloyd) complexity, which is expected to peak for structured rather than random states. The CA prescription, by contrast, yields a negative correction relative to the vacuum and relaxes back to its unperturbed value at late times, both in tension with standard expectations for complexity of formation. Their conclusion is that CV and CA capture different aspects of physical complexity.  The disagreement between the spread complexity of the quenches studied in our work and the proposed momentum of the falling particle in the bulk is in the direction of the broader question of how different boundary notions of complexity are encoded holographically. Hence, it will be an interesting subject to scrutinize further and identify the relevant bulk observable to the spread/Krylov complexity.

Several further directions are natural. It will be very interesting to obtain a closed form for the Lanczos coefficients derived here to find an analytic expression for the spread complexity of the states. As mentioned before, it is also desired to find a holographic dual to the spread complexity to match the results calculated on the CFT side.

Another line of research is studying states excited by operators that correspond to extended objects in the bulk of the dual theory. The spread complexity associated with these states will be of great interest to compare with different proposed holographic observables that are present in the literature \cite{Nastase:2026lhz,Chatzis:2026ekd,Chatzis:2026oou}. Also, making connections with studies of variation of initial state and its effect on the spread complexity is desired \cite{Balasubramanian:2025xkj}.

We hope that the results obtained here can be useful as a benchmark for future holographic, lattice, and QFT studies of spread complexity for states with conserved momentum.

\section*{Acknowledgments}

We would like to thank Felipe Diaz, Horatiu Nastase, and Carlos Nunez for sharing their ideas and useful discussions. We are thankful to Horatiu Nastase and Carlos Nunez for constructive comments on the manuscript. Standard AI programming tools aided the development of the code used to produce the results in this paper, primarily improving efficiency in debugging and code modification.

\vspace{0.5cm}

{\bf Open Access Statement} --- For the purpose of open access, the authors have applied a Creative Commons Attribution (CC BY) license to any Author Accepted Manuscript version arising. 

\vspace{0.5cm}

{\bf Research Data Access Statement} --- The data generated for this manuscript can be downloaded from \cite{fatemiabhari_2}.

\vspace{0.5cm}

\hrule

\appendix
\section{Toda-chain method} \label{app:Lanc1}
In this appendix, we provide the prescription for the Toda-chain method for computing the Lanczos coefficients $a_n$ and $b_n$. A more detailed derivation can be found in \cite{Dymarsky:2019elm,Kundu:2023hbk}.
Consider the so-called Toda tau functions $\tau_n(\tau)$, which obey the Hirota bilinear equations
\begin{align} \label{eq:toda}
	\tau_n \ddot{\tau}_n-\dot{\tau}_n^2=\tau_{n+1} \tau_{n-1}\,,
\end{align}
where dots denote derivatives with respect to $\tau$, the Euclidean time. The chain is closed by
the boundary conditions
\begin{align}
\tau_{-1}\equiv 1\,,\qquad \tau_{0}(\tau)\equiv S(\tau)\,,
\end{align}
with $S(\tau)$ the Euclidean autocorrelation function of the state. $S(\tau)$ can be obtained from $S(t)$ in Eq.~\eqref{eq:Stdef} by $t\to -i\tau$. Given
these, Eq.~\eqref{eq:toda} provides all  $\tau_n$ functions algebraically through
\begin{align}
\tau_{n+1}=\frac{\tau_n\ddot{\tau}_n-\dot{\tau}_n^2}{\tau_{n-1}}\,,
\end{align}
without any differential equation to integrate; equivalently, one can obtain $\tau_n$ through
$(n+1)\times(n+1)$ Hankel determinant,
\begin{equation} \label{eq:Hankel}
\tau_n=\det\bigl[\partial_\tau^{\,i+j}S(\tau)\bigr]_{i,j=0}^{n}.
\end{equation}
The Lanczos coefficients then follow as \cite{Dymarsky:2019elm}
\begin{align}
b_{n+1}^2&=\left.\frac{\tau_{n-1} \tau_{n+1}}{\tau_n^2}\right|_{\tau=0}\,,\label{eq:bntoda} \\
a_n&=\left.\frac{d}{d \tau}\log\frac{\tau_n}{\tau_{n-1}}\right|_{\tau=0}\,. \label{eq:antoda}
\end{align}
Here we stick to the convention $b_0=0$.

\section{Lancsoz coefficients} \label{app:Lanc}
In this appendix, we mention the difficulties of obtaining the Lanczos coefficients through the methods described in Appendix~\ref{app:Lanc1}. Then we will provide a method that can be applied numerically to obtain the Lanczos coefficients up to higher orders in the Krylov index $n$ and then derive the spread/Krylov complexity associated with them.

It is a well-known fact that computing Lanczos coefficients numerically through the methods in Appendix~\ref{app:Lanc1} is ill-conditioned. The Hankel matrix whose determinant is taken in Eq.~\eqref{eq:Hankel} has some elements that are of order one and many elements that have very small, nearly degenerate values. Hence, calculation of $b_n$ and $a_n$ in Eqs.\eqref{eq:bntoda}-\eqref{eq:antoda} suffers from rounding errors in subtraction and division of floating point numbers, and the process is inherently sensitive to small perturbations and errors. This is closely related to the ill-defined Hilbert matrix problem.

The core issue lies in the nonlinear map from the moments of a distribution to the Lanczos coefficients as discussed in \cite{Gautschi82}. There, it is proved that small relative errors in the input moments can produce exponentially large relative errors in the computed coefficients, and the sensitivity is not a flaw of a particular algorithm but a fundamental property of the problem itself.
The way to overcome is to bypass the process of $S(t)\to{\mu_i}\to\{a_n,b_n\}$ and find another relation between return amplitude and the coefficients.  In the next section, we will review one.

\subsection{The spectral method}
In order to make this section self-contained, we repeat some definitions. Let ${\cal H}$ be the relevant Hilbert space and $|\psi_0\rangle$ the initial state, evolving as
\begin{equation}
|\psi(t)\rangle = e^{-iHt}|\psi_0\rangle.
\end{equation}
The return amplitude can be written in the inverse Fourier form as
\begin{equation} \label{eq:measure}
S(t)^*=\langle \psi_0|\psi(t)\rangle=\int_{\mathbb R} e^{-i\omega t}\,d\mu(\omega),
\end{equation}
where $d\mu$ is the spectral measure of $H$ in the state $|\psi_0\rangle$,
\begin{equation}
d\mu(\omega)=d\langle \psi_0|P_\omega|\psi_0\rangle.
\end{equation}
Here $P_\omega$ is the so-called projection-valued measure. This is a one-parameter family of orthogonal projectors, indexed by a real number $\omega$, and its definition will become clear in the discrete spectrum example that follows. 
Suppose $H$ has a discrete non-degenerate spectrum,
\begin{equation}
H |E_k\rangle = E_k |E_k\rangle .
\end{equation}
Then
\begin{equation}
S(t)^*
= \langle \psi_0 | e^{-iHt} | \psi_0 \rangle
= \sum_k |\langle E_k | \psi_0 \rangle|^2 e^{-iE_k t}.
\end{equation}
If we define $p_k = |\langle E_k | \psi_0 \rangle|^2$, then $S(t)^*$ is the Fourier transform of the discrete measure
\begin{equation}
d\mu(\omega)
= \sum_k p_k \, \delta(\omega - E_k) \, d\omega .
\end{equation}
Hence $\mu$ is the probability distribution of the energy in the state $|\psi_0\rangle$ and $S(t)$ is its Fourier transform. In the discrete case,
\begin{equation}
P_\omega
= \sum_{k:\, E_k \le \omega} |E_k\rangle \langle E_k|,
\end{equation}
is the projector onto the span of all eigenvectors with energy at or below $\omega$. As $\omega$ runs from $-\infty$ to $+\infty$, $E_\omega$ climbs in steps from $0$ to $1$. Each time $\omega$ crosses an eigenvalue $P_\omega$, jumps by $|E_k\rangle \langle E_k|$ .

Now we move on from the discrete example and continue with the definition of the moments that are given as
\begin{equation}
\mu_n=\int \omega^n\,d\mu
=\langle \psi_0|H^n|\psi_0\rangle.
%=i^n\left.\frac{d^n S}{dt^n}\right|_{t=0}.
\end{equation}

\subsubsection{The Krylov space and its isomorphism with $L^2(d\mu)$}
As explained in Section~\ref{sec:spread}, the Krylov space is given by 
$\mathcal K= \overline{\operatorname{span}}\{ H^n |\psi_0\rangle \}_{n\ge 0}$
and Gram-Schmidt orthogonalisation on
\begin{equation}
\{ |\psi_0\rangle,\ H|\psi_0\rangle,\ H^2|\psi_0\rangle,\ \dots \}
\end{equation}
yields the orthonormal Krylov basis $\{ |K_n\rangle \}$. Since $|K_n\rangle$ is by construction a degree-$n$ polynomial in ${\cal H}$ acting on $|\psi_0\rangle$, we may write
\begin{equation}
|K_n\rangle = \frac{ p_n(H)}{\sqrt{h_n}}|\psi_0\rangle=\hat p_n(H)|\psi_0\rangle, \qquad
h_n=\int p_n^2(\omega)\,d\mu,
\end{equation}
with $p_n(H)$ being monic by definition so the coefficient of the highest-degree term is one.
One can show that the map --- see also Refs.~\cite{golub2010matrices,Muck:2022xfc,Nandy:2024evd}
\begin{equation}
|K_n\rangle
\longleftrightarrow
\hat p_n(\omega)=\frac{p_n(\omega)}{\sqrt{h_n}},
% \qquad
% h_n=\int p_n^2(\omega)\,d\mu
\end{equation}
is a unitary isomorphism
\begin{equation}
\mathcal K \to L^2(\mathbb R,d\mu),
\end{equation}
where $L^2(\mathbb R,d\mu)$ is the space of square-integrable functions on the real line with respect to the measure $d\mu$.
Under this map $H$ becomes multiplication by $\omega$. To see these define
\begin{equation}
U : L^2(\mathbb R,d\mu) \longrightarrow K,
\qquad
U f = f(H) |\psi_0\rangle,
\end{equation}
using the functional calculus $f(H)=\int f(\omega)\, dP_\omega.$
Taking two polynomials $f,g$ one can compute the inner product of their images as
\begin{equation}
\langle f(H)\psi_0 \mid g(H)\psi_0 \rangle
= \langle \psi_0 \mid \overline{f}(H) g(H) \mid \psi_0 \rangle
= \langle \psi_0 \mid (\overline{f}g)(H) \mid \psi_0 \rangle
= \int \overline{f(\omega)}\, g(\omega)\, d\mu(\omega).
\end{equation}
So $U$ preserves inner products and it is an isometry, and $L^2(d\mu)$ is a faithful copy of the Krylov space.

Let $m_\omega$ denote ``multiply by $\omega$'' action,
\begin{equation}
(m_\omega f)(\omega)=\omega f(\omega).
\end{equation}
Then
\begin{equation}
U(m_\omega f)
= (\omega f)(H) |\psi_0\rangle
= H f(H) |\psi_0\rangle
= H (U f).
\end{equation}
Hence, multiplication by $\omega$ in $L^2(d\mu)$ is equivalent to acting by $H$ on states.

Also, orthonormality of the Krylov basis is exactly orthonormality of the polynomials:
\begin{equation}
\langle K_n|K_m\rangle
= \int \overline{\hat p_n(\omega)}\,\hat p_m(\omega)\,d\mu(\omega)
= \delta_{nm}.
\end{equation}

An important point is that $p_n(\omega)$ are monic by construction, but not normalised, and $\hat p_n(\omega)$ are the normalised versions.

\textbf{Monic recurrence:} Now we derive a recursion relation for the $p_n$ polynomials. Multiplication by $\omega$ is symmetric on $L^2(d\mu)$:
\begin{equation} \label{eq:symAd}
(\omega f,g)=(f,\omega g).
\end{equation}
Hence, for expanding $\omega p_n$, which is a polynomial of degree $n+1$, in the orthogonal basis the coefficient of $p_m$ obeys $(\omega p_n,p_m) =(p_n,\omega p_m),$ which vanishes whenever $\deg(\omega p_m)=m+1<n$, i.e. for all $m<n-1$. Only $m=n-1,n,n+1$ survive in the \textit{monic recurrence}:
\begin{equation} \label{eq:monic}
\omega p_n(\omega)
=
p_{n+1}(\omega)+a_n p_n(\omega)+b_n^2 p_{n-1}(\omega).
\end{equation}
The coefficients are derived in the following way. We pair the monic recurrence with $p_n$
\begin{equation}
(\omega p_n, p_n) = a_n h_n
\quad\Longrightarrow\quad
a_n = \frac{(\omega p_n, p_n)}{h_n}
= \frac{1}{h_n} \int \omega\, p_n(\omega)^2 \, d\mu.
\end{equation}
If we pair it with $p_{n-1}$:
\begin{equation}
(\omega p_n, p_{n-1}) = b_n^2 \, h_{n-1}= (p_n, \omega p_{n-1}).
\end{equation}
The last relation uses Eq.~\eqref{eq:symAd}. The fact that $p_n$s are monic leads to 
$\omega p_{n-1} = p_n + (\text{degree} \le n-1),$
so this equals
\begin{equation}
(p_n, \omega p_{n-1})=(p_n, p_n) = h_n.
\end{equation}
Finally 
\begin{equation}
b_n^2 = \frac{h_n}{h_{n-1}}.
\end{equation}
This is why the coefficient of $p_{n-1}$ in Eq.~\eqref{eq:monic} is positive and is defined as $b_n^2 \ge 0$.
%%%%

If we divide the monic form in Eq.~\eqref{eq:monic} by $\sqrt{h_n}$,
with $b_n^2=h_n/h_{n-1}$, the two off-diagonal coefficients become $b_{n+1}$ and $b_n^2/b_n=b_n$:
\begin{equation} \label{eq:recursion}
\omega \hat p_n
=
b_{n+1}\hat p_{n+1}+a_n\hat p_n+b_n\hat p_{n-1}.
\end{equation}
This is equivalent to the discrete Schrodinger equation in Eq.~\eqref{eq:schrEq}. Indeed Applying the isometry to
$\varphi_n(t)=\langle K_n | e^{-iHt} | \psi_0 \rangle$ leads to
\begin{equation}
\varphi_n(t)
= \int \hat p_n(\omega)\, e^{-i\omega t}\, d\mu(\omega)
= \frac{1}{\sqrt{h_n}} \int p_n(\omega) e^{-i\omega t} d\mu(\omega).
\end{equation}
For $n=0$, $p_0=1$, $h_0=\int d\mu=1$, hence
\begin{equation}
\varphi_0(t)=\int e^{-i\omega t} d\mu = S(t)^*,
\end{equation}
and the return amplitude is the complex conjugate to the zeroth Krylov wavefunction.

\subsubsection{Numerical construction of coefficients via discretisation}
Now we present a method that bypasses the moments of return amplitude to derive the Lanczos coefficients. 
The solution is approximating $\mu$ by a discrete measure
\begin{equation}
\mu_M = \sum_{i=1}^{M} c_i \, \delta_{\omega_i}
\end{equation}
This is exact, if $\mu$ is discrete (atomic), or is an approximation if it is continuous, and the best choice of discretisation points on the $\omega$ line will be by Gauss quadrature. Note that the $\mu_M$ here should not be confused with moments of return amplitude $\mu_i$. 

After discretisation, one can run a discretised version of the Lanczos recursion in Eq.~\eqref{eq:recursion} 
\begin{equation}
r_n = \Omega u_n - a_n u_n - b_n u_{n-1},
\qquad
a_n = u_n^\top \Omega u_n,
\qquad
b_{n+1} = \| r_n \|,
\qquad
u_{n+1} = r_n / b_{n+1}.
\end{equation}
where $\Omega=\operatorname{diag}(\omega_i)$, with $v_i = \sqrt{c_i}$ and $u_0 = \sqrt{h_0}\, v = \frac{v}{\|v\|}$. The vectors $(u_n)_i$ are equivalent to $c_i p_n(\omega_i)$.

This is the Gragg-Harrod or Rutishauser-Kahan-Pal-Walker (RKPW) algorithm for reconstructing a Jacobi matrix (Hamiltonian) from spectral data \cite{Gragg1984}, and it is numerically stable where the moment map is not. The reason is the following. $p_n(\omega)$ oscillates $n$ times, so in the monomial basis $\{\omega^n\}$ its coefficients alternate in sign with magnitudes vastly exceeding the values of $p_n$ itself. The discretisation changes the representation of the polynomial, and we never store $p_n$ by coefficients in $\{\omega^n\}$ but by its values on the quadrature nodes $(u_n)_i=c_i p^n(\omega_i)$. This way, the inner products become simple Euclidean dot products of O(1) sized vectors with no cancellation due to large alternating coefficients in $\{\omega^n\}$ basis. Hence, the algorithm provides the coefficients $a_n$ and $b_n$ stably through recursion. 

Caution needs to be practised in choosing the grid size $M$ and the range of the grid. A grid of size $M$ can at most accommodate polynomials of degree $M$; hence, in order to find $a_n$ and $b_n$ up to high orders $n_{max}=N$, which requires finding $p_N(\omega)$, one has to choose $M>>N$. Second point is that the grid must span an interval in $\omega$ that carries the support of $|\hat p_N|^2 \, d\mu$ appropriately.

\subsubsection{Complexity calculation}
Having $a_n$ and $b_n$ at hand up to order $N$ we have the truncated Jacobi (Hamiltonian) matrix $H_N$, defined as
\begin{equation}
H_N =
\begin{pmatrix}
a_0    & b_1    &        &        &        \\
b_1    & a_1    & b_2    &        &        \\
       & b_2    & a_2    & \ddots &        \\
       &        & \ddots & \ddots & b_N    \\
       &        &        & b_N    & a_N
\end{pmatrix}
\end{equation}
It is real and symmetric, so one can find a real orthogonal eigenbasis for it:
\begin{equation}
H_N |v_k\rangle = E_k |v_k\rangle,
\qquad
k = 0,\dots,N,
\qquad
\langle v_k | v_l \rangle = \delta_{kl}.
\end{equation}
Define $U$ as the matrix whose columns are the eigenvectors,
\begin{equation}
U_{nk} = (v_k)_n = \langle n | v_k \rangle.
\end{equation}
% So the row index $n$ labels the Krylov site, the column index $k$ labels the eigenvector. Then $U^\dagger U = U U^\dagger = \mathbb{1}$ and $H_N = U \operatorname{diag}(E_k) U^\dagger$, i.e.
% \begin{equation}
% H_N = \sum_k E_k |v_k\rangle \langle v_k|.
% \end{equation}
Basically, $U_{nk}$ is the amplitude of Krylov site $n$ in the $k$-th eigenvector, and $U_{0k}$ is that eigenvector's overlap with the initial state $|K_0\rangle$.

The Krylov wavefunction is
\begin{equation}
\varphi_n(t) = \langle K_n | e^{-iHt} | K_0 \rangle\simeq \langle n | e^{-iH_N t} | 0 \rangle.
\end{equation}
Hence,
\begin{equation}
\varphi_n(t)
= \sum_k \langle n | v_k \rangle \, e^{-iE_k t} \, \langle v_k | 0 \rangle
= \sum_k U_{nk} \, U_{0k} \, e^{-iE_k t}.
\end{equation}
So by diagonalising the $H_N$ matrix and obtaining $U_{nk}$, one can easily calculate $C(t) = \sum_{n=0}^{N} n \, |\varphi_n(t)|^2$. This completes the procedure to obtain $C(t)$ from the return amplitude $S(t)$.

\subsubsection{Fourier transform of return amplitudes}
The return amplitudes studied in this work have exact Fourier transforms that we make use of for the calculation of the Lanczos coefficients. Here we report the transformed functions. We report them in the convention
\begin{equation}
\Phi(\omega) = \int_{-\infty}^{\infty} f(t) e^{-i\omega t} \, dt,
\end{equation}
that can be transformed to other conventions as they are needed.

\textbf{CFT on the plane:} For $S(t)$ given in Eq.~\eqref{eq:StLin},
\begin{equation}
S(t)=\left(1-\frac{it}{2\epsilon_+}\right)^{-\Delta}\left(1-\frac{it}{2\epsilon_-}\right)^{-\Delta},
\end{equation}
the fourier transform is \cite{bateman1954tables}
\begin{equation}
\Phi(\omega)
=\Theta(\omega)\,\frac{2\pi (4\epsilon_- \epsilon_+)^\Delta}{\Gamma(2\Delta)}\,\omega^{2\Delta-1}\,e^{-2\epsilon_+ \omega}\,{}_1 F_1\!\left(\Delta; 2\Delta; 2(\epsilon_+-\epsilon_-)\omega\right)
\end{equation}
where $\Theta(\omega)$ is the Heaviside step function and ${}_1 F_1(a;b;z)$ is the confluent hypergeometric function of the first kind.

\textbf{CFT on the cylinder:} For $S(t)$ given in Eq.~\eqref{eq:CylSt},
\begin{equation}
S(t)=\left(\frac{\sinh\left(\frac{2\pi\epsilon_+}{L}\left(1-\frac{it}{2\epsilon_+}\right)\right)}{\sinh\left(\frac{2\pi\epsilon_+}{L}\right)}\right)^{-\Delta}\left(\frac{\sinh\left(\frac{2\pi\epsilon_-}{L}\left(1-\frac{it}{2\epsilon_-}\right)\right)}{\sinh\left(\frac{2\pi\epsilon_-}{L}\right)}\right)^{-\Delta},
\end{equation}
the Fourier transform is 
\begin{equation}
\Phi(\omega) = 2\pi \sum_{n=0}^{\infty} A_n \, \delta\left(\omega - \frac{2\pi(n+\Delta)}{L}\right),
\end{equation}
with
\begin{equation}
A_n = (1-q_+)^\Delta (1-q_-)^\Delta \sum_{k=0}^{n} \frac{(\Delta)_k (\Delta)_{n-k}}{k!(n-k)!} q_+^k q_-^{n-k}.
%= (1-q_+)^\Delta (1-q_-)^\Delta \frac{(\Delta)_n}{n!} q_-^n \, {}_2 F_1\left(-n, \Delta; 1-\Delta-n; \frac{q_+}{q_-}\right).
\end{equation}
Here $q_\pm = e^{-4\pi \epsilon_\pm / L}$ .
Since $S(t)$ is periodic in this case, the Fourier transform is a Dirac comb and makes the discretisation trivial.

\textbf{CFT at finite temperature:} For $S(t)$ given in Eq.~\eqref{eq:StThe},
\begin{equation}
S(t)=\left(\frac{\sin\left(\frac{2\pi\epsilon_+}{\beta}\left(1-\frac{it}{2\epsilon_+}\right)\right)}{\sin\left(\frac{2\pi\epsilon_+}{\beta}\right)}\right)^{-\Delta}\left(\frac{\sin\left(\frac{2\pi\epsilon_-}{\beta}\left(1-\frac{it}{2\epsilon_-}\right)\right)}{\sin\left(\frac{2\pi\epsilon_-}{\beta}\right)}\right)^{-\Delta},
\end{equation}
The Fourier transform is \cite{bateman1954tables}
\begin{equation}
\Phi(\omega)
=\left[(\zeta_--1)(\zeta_+-1)\right]^\Delta\,\frac{\beta}{2\pi}\,(-\zeta_-)^{-s}\,\frac{\Gamma(s)\Gamma(2\Delta-s)}{\Gamma(2\Delta)}\;{}_2 F_1(\Delta,s;2\Delta;z),
\end{equation}
with
$s=\Delta+i\beta\omega/2\pi$, $z=1-\zeta_+/\zeta_-$, and $\zeta_j = e^{4\pi i \varepsilon_j/\beta}.$ The function ${}_2 F_1(a,b;c;z)$ is the Gaussian hypergeometric function.

\section{Gravity duals} \label{app:gravity}
In this appendix, we review how the back-reacted geometries of infalling particles with conserved momentum were obtained in \cite{Caputa:2026hxt} by boosting some static background solutions.

We consider geometries that are asymptotically $AdS_3$ with radius $R$. 
For $\mathbb{R}^{2,2}$ with metric
\begin{equation}
ds^2=-dX^2_0-dX^2_1+dX^2_2+dX^2_3\,,
\end{equation}
pure $AdS_3$ is the hyperboloid
\begin{eqnarray}
-X^2_0-X^2_1+X^2_2+X^2_3=-R^2\,.\label{eq:embedding}
\end{eqnarray}
Different parametrizations of Eq.~\eqref{eq:embedding} yield the global,
Poincare and thermal coordinate patches were used previously in the article. In a chosen patch, we place a massive point particle on a timelike geodesic that starts near the asymptotic boundary with vanishing radial velocity but non-zero spatial or angular momentum.

The back-reaction of such a particle is constructed by the
following procedure \cite{Horowitz:1999gf,Nozaki:2013wia,Caputa:2026hxt}. The $SO(2,2)$ isometry group of $AdS_3$ contains two commuting boosts, with rapidities
$\eta_1$ and $\eta_2$,
\begin{eqnarray}
\tilde{X}_0&=&X_0\cosh\eta_1-X_3\sinh\eta_1\,,\nn\\
\tilde{X}_1&=&X_1\cosh\eta_2-X_2\sinh\eta_2\,,\nn\\
\tilde{X}_2&=&X_2\cosh\eta_2-X_1\sinh\eta_2\,,\nn\\
\tilde{X}_3&=&X_3\cosh\eta_1-X_0\sinh\eta_1\,.\label{eq:mGenral}
\end{eqnarray}
For an appropriate choice of $(\eta_1,\eta_2)$, the map sends the moving particle in the original  $X_i$ coordinates to a particle at rest at the origin of global $AdS_3$ parametrised by the coordinates $\tilde{X}_i$. This requirement fixes the boost parameters in terms of the physical data of the original configuration, namely the initial radial position and the spatial or angular momentum of the particle.

In the frame where the particle is static, its back-reaction is known in closed form,
\begin{equation}
ds^2=-(r^2+R^2-M)\,d\tau^2+\frac{R^2\,dr^2}{r^2+R^2-M}+r^2d\theta^2\,,\label{eq:staticMetric}
\end{equation}
which describes a conical defect for $0<M<R^2$ and a BTZ black hole for
$M>R^2$. The particle mass is given in terms of $M$ by
\begin{equation}
m=\frac{M}{8G_NR^2}\,,
\end{equation}
with $G_N$ the three-dimensional Newton constant. 

As Eq.~\eqref{eq:staticMetric} is locally $AdS_3$ everywhere away from the particle, the inverse of Eq.~\eqref{eq:mGenral} is applicable and maps the static solution back to the original coordinates. This yields the exact back-reacted geometry of the moving particle as an asymptotically $AdS_3$ spacetime.
Now we consider different parametrisations of the embeddings.

\textbf{Poincare AdS:}
If one considers 
\begin{eqnarray}
\sqrt{r^2+R^2}\cos(\tau)&=&\tilde{X}_0=\frac{e^{-\eta_1}(z^2+x^2-t^2)+R^2e^{\eta_1}}{2z}\,,\nn\\
\sqrt{r^2+R^2}\sin(\tau)&=&\tilde{X}_1=R\frac{t\cosh(\eta_2)-x \sinh(\eta_2)}{z}\,,\nn\\
r\sin(\theta)&=&\tilde{X}_2=R\frac{x\cosh(\eta_2)-t \sinh(\eta_2)}{z}\,,\nn\\
r\cos(\theta)&=&\tilde{X}_3=\frac{e^{-\eta_1}(z^2+x^2-t^2)-R^2e^{\eta_1}}{2z}\,,\label{eq:mPoincare}
\end{eqnarray}
one can obtain the $AdS_3$ metric in the Poincare coordinates given as
\begin{equation}
ds^2=R^2\frac{-dt^2+dx^2+dz^2}{z^2}\,.
\end{equation}
The metric, before applying boosts, describing the particle at rest at the origin $r=0$ is
\begin{equation}
ds^2=-(r^2+R^2)d\tau^2+\frac{R^2dr^2}{r^2+R^2}+r^2d\theta^2\,.
\end{equation}
The identifications
\begin{equation}
Re^{\eta_1}=\epsilon,\qquad \tanh(\eta_2)=v\,,
\end{equation}
relates the boost parameters to the initial value data of the infalling particle defined in Eq.~\eqref{eq:IVPoincare}. Then one can match the holographic stress tensor calculation results from this back-reacted geometry to the CFT result quoted in Eq.~\eqref{TpmPoincare}.

\textbf{Global AdS:}
For a similar analysis in global AdS, one starts with the map Eq.~\eqref{eq:mGenral} with global coordinates 
\begin{eqnarray}
\sqrt{r^2+R^2}\cos(\tau)&=&X_0=R\cosh(\rho)\cos\left(\frac{2\pi t}{L}\right)\,,\nn\\
\sqrt{r^2+R^2}\sin(\tau)&=&X_1=R\cosh(\rho)\sin\left(\frac{2\pi t}{L}\right)\,, \\ 
r\sin(\theta)&=&X_2=R\sinh(\rho)\sin\left(\frac{2\pi \phi}{L}\right)\,,\nn\\
r\cos(\theta)&=&X_3=R\sinh(\rho)\cos\left(\frac{2\pi \phi}{L}\right)\,.\label{eq:mGlobal}
\end{eqnarray}
The relations on the left are the same as Eq.~\eqref{eq:mPoincare} while the relations on the right-hand side should be inserted into Eq.~\eqref{eq:mGenral} to generate the boosted back-reacted metric. 

The identifications
\begin{align}
\rho_\Lambda\,&=\eta_1,\\
\omega&=\frac{\tanh(\eta_2)}{\tanh(\eta_1)}\,,    
\end{align}
relate the boost parameters to the initial value data of the infalling particle. Then one can match the holographic stress tensor calculation results from this back-reacted geometry to the CFT result quoted in Eq.~\eqref{TpmGlobal}.

The map for the BTZ black hole can similarly be obtained, and we refer the reader to the original works in Refs.~\cite{David:2026owc,Caputa:2026hxt}.

\bibliographystyle{JHEP}
% \textcolor{red}{Corrected bibliography command: BibTeX expects the database name without the .bib extension.}
\bibliography{main}

\end{document}